\documentclass[10pt,journal]{IEEEtran}
\usepackage[ruled,linesnumbered]{algorithm2e}
\usepackage{amsthm}
\usepackage{amsmath}
\usepackage{graphicx} 
\usepackage{epsfig,graphics,psfrag,amsmath,amssymb}
\usepackage{cite}
\usepackage{float}
\usepackage{tocvsec2}
\usepackage{enumitem}
\usepackage{color}
\usepackage{verbatim}
\usepackage{mathtools}
\usepackage{hhline}
\usepackage{enumerate}
\usepackage{arydshln}
\usepackage{bbm}
\usepackage{stackengine}
\usepackage{mathrsfs}
\usepackage{enumitem}
\usepackage{caption}
\usepackage{stfloats}
\usepackage{amssymb}
\usepackage{hyperref}       % hyperlinks
\usepackage[export]{adjustbox} 

\usepackage{setspace}

\usepackage[symbol]{footmisc}
\renewcommand{\thefootnote}{\fnsymbol{footnote}}
\newcommand \footnoteONLYtext[1]
{
	\let \mybackup \thefootnote
	\let \thefootnote \relax
	\footnotetext{#1}
	\let \thefootnote \mybackup
	\let \mybackup \imareallyundefinedcommand
}

\newtheorem{thm}{Theorem}
\newtheorem{lem}{Lemma}

\newtheorem{defi}{Definition}
\newtheorem{rem}{Remark}

\makeatletter

\newcommand{\Rmnum}[1]{\expandafter\@slowromancap\romannumeral #1@}
\makeatother

\makeatletter
\newcommand*\bigcdot{\mathpalette\bigcdot@{.5}}
\newcommand*\bigcdot@[2]{\mathbin{\vcenter{\hbox{\scalebox{#2}{$\m@th#1\bullet$}}}}}
\makeatother

\def\@#1{\pmb{#1}}
\def\b#1{\mathbb{#1}}
\def\bf#1{\mathbf{#1}}
\def\s#1{\mathsf{#1}}

\def\ca#1{\mathcal{#1}}

\newcommand\aleq{\mathrel{\stackrel{\makebox[0pt]{\mbox{\normalfont\tiny (a)}}}{\leq}}}

\newcommand\aeq{\mathrel{\stackrel{\makebox[0pt]{\mbox{\normalfont\tiny (a)}}}{=}}}

\usepackage{makecell}

\usepackage{amsmath,graphicx}
\usepackage{amsmath,amssymb,amsfonts}
\usepackage{cite, soul}
\usepackage{url}
\usepackage{amsthm}
\usepackage{bm}
\usepackage{comment}
\usepackage{multirow}
\usepackage{booktabs}
\usepackage{hhline}
\usepackage{graphicx}     % 插入图片
\usepackage{caption}      % 自定义标题
\usepackage{subcaption}   % 支持子图

\usepackage{setspace}
\usepackage{setspace}
\usepackage[ruled,linesnumbered]{algorithm2e}

\def\@#1{\pmb{#1}}
\def\b#1{\mathbb{#1}}
\def\bf#1{\mathbf{#1}}
\def\s#1{\mathsf{#1}}

\def\ca#1{\mathcal{#1}}

\makeatletter % 重点代码区开始 <=======================================================
\let\myorg@bibitem\bibitem
\def\bibitem#1#2\par{%
	\@ifundefined{bibitem@#1}{%
		\myorg@bibitem{#1}#2\par
	}{%
		\begingroup
		\color{\csname bibitem@#1\endcsname}%
		\myorg@bibitem{#1}#2\par
		\endgroup
	}%
}
\newcommand*{\bibitem@fanjoint}{black}
\newcommand*{\bibitem@liangcommunicationaaa}{black}
\newcommand*{\bibitem@sunfederated}{black}
\newcommand*{\bibitem@zhuangintegrated}{black}
\newcommand*{\bibitem@yangover}{black}
\newcommand*{\bibitem@kaiicc}{black}
\newcommand*{\bibitem@dllama}{black}
\newcommand*{\bibitem@liangcommunicationaaaf}{black}
\newcommand*{\bibitem@liangcommunicationaaag}{black}
\makeatother

\title{Communication-Efficient Distributed On-Device LLM Inference Over Wireless Networks}
\author{Kai Zhang, Hengtao He, \emph{Member}, \emph{IEEE}, Shenghui Song, \emph{Senior Member}, \emph{IEEE},\\Jun Zhang, \emph{Fellow}, \emph{IEEE}, and Khaled B. Letaief, \emph{Fellow}, \emph{IEEE}
}

\begin{document}

%\vspace{-40pt}

%\input{figure.tex}
\maketitle

\footnoteONLYtext{
	{\color{black} \hspace*{-1.599em} 
	Part of this work has been accepted for presentation at the 2025 \emph{IEEE Int. Conf. Commun. (ICC)}, Montreal, Canada \cite{kaiicc}.}
	
	The authors are with the Department of Electronic and Computer Engineering, The Hong Kong University of Science and Technology, Clear Water Bay, Hong Kong (email: kzhangbn@connect.ust.hk, eehthe@ust.hk, eeshsong@ust.hk, eejzhang@ust.hk, eekhaled@ust.hk). (The corresponding author is Hengtao He.)
}

\begin{abstract}
	
Large language models (LLMs) have demonstrated remarkable success across various application domains, but their enormous sizes and computational demands pose significant challenges for deployment on resource-constrained edge devices. To address this issue, we propose a novel distributed on-device LLM inference framework that leverages tensor parallelism to partition the neural network tensors (e.g., weight matrices) of one LLM across multiple edge devices for collaborative inference. A key challenge in tensor parallelism is the frequent all-reduce operations for aggregating intermediate layer outputs across participating devices, which incurs significant communication overhead. To alleviate this bottleneck, we propose an over-the-air computation (AirComp) approach that harnesses the analog superposition property of wireless multiple-access channels to perform fast all-reduce steps. To utilize the heterogeneous computational capabilities of edge devices and mitigate communication distortions, we investigate a joint model assignment and transceiver optimization problem to minimize the average transmission error. The resulting mixed-timescale stochastic non-convex optimization problem is intractable, and we propose an efficient two-stage algorithm to solve it. Moreover, we prove that the proposed algorithm converges almost surely to a stationary point of the original problem. Comprehensive simulation results will show that the proposed framework outperforms existing benchmark schemes, achieving up to 5x inference speed acceleration and improving inference accuracy.

\end{abstract}

\begin{IEEEkeywords} 6G, distributed inference, large language models, over-the-air computation, tensor parallelism. \end{IEEEkeywords} 

\section{Introduction}

The advent of large language models (LLMs) has marked a significant breakthrough in artifical intelligence (AI), demonstrating superior performance and adaptability in a wide range of applications, such as natural language processing \cite{min2023recent,chang2024survey,gu2021domain}, embodied intelligence \cite{fan2024embodied,yang2024embodied,driess2023palm}, and wireless communications \cite{bariah2024large,shao2024wirelessllm,tong2024wirelessagent}.
The efficacy of LLMs is primarily attributed to the vast model scale with billions of parameters, which enables them to capture complex semantic relationships and contextual nuances, leading to superior performance across diverse tasks.
However, the substantial computational and memory requirements of LLMs present significant challenges for the deployment on resource-constrained edge devices.
For instance, the LLaMA3 model \cite{dubey2024llama} with 13 billion parameters requires 40GB of RAM, which far exceeds the capabilities of most edge devices.
Consequently, most existing LLMs rely on cloud-based infrastructure, which limits the feasibility of LLM deployment and raises concerns about data privacy and inference latency, especially in sensitive domains like healthcare and finance.
To address these challenges, distributed LLM inference has recently been proposed as a promising solution, which distributes the large models and computational workloads across multiple devices \cite{wu2023fast,borzunov2024distributed, hu2024inference}. This strategy allows each device to handle smaller and more manageable model segments, thereby reducing the burden on individual devices and strengthening privacy protections.
Furthermore, advancements in communication technologies, such as the 5G and future 6G wireless networks, enhance the feasibility of distributed LLM inference for real-time applications \cite{letaief2019roadmap,letaief2021edge}.

Communication overhead is a critical factor affecting the performance of distributed LLM inference systems.
To enhance communication efficiency, several recent studies have been conducted \cite{zhang2024edgeshard, yuan2024high, he2024large, chen2024adaptive, shao2021learning,li2023task,li2024tackling,wen2023task}.
In \cite{zhang2024edgeshard}, Zhang \emph{et al.} proposed a collaborative edge computing framework that distributes different layers of LLMs across the edge device and cloud server. They developed a joint device selection and model partitioning algorithm to minimize inference latency and maximize throughput.
In \cite{yuan2024high}, Yuan \emph{et al.} considered splitting LLMs into several sub-models, where the resource-intensive components were offloaded to the server through non-orthogonal multiple-access (NOMA) channels. They further proposed a gradient descent-based algorithm to find the optimal trade-off between inference delay and energy consumption.
In \cite{he2024large}, He \emph{et al.} developed an active inference method to address the joint task offloading and resource allocation problem for distributed LLM inference over cloud-edge computing frameworks.
Similarly, Chen \emph{et al.} \cite{chen2024adaptive} proposed a reinforcement learning algorithm that optimizes the splitting point of LLMs between the edge device and cloud server to reduce the communication overhead under varying wireless network conditions.
Furthermore, task-oriented communications have been utilized to optimize end-to-end inference throughput, accuracy, and latency, which can further enhance the communication efficiency of distributed LLM inference systems \cite{shao2021learning,li2023task,li2024tackling,wen2023task}.

Despite significant advances in distributed LLM inference, most existing works \cite{zhang2024edgeshard, yuan2024high,chen2024adaptive,he2024large, shao2021learning,li2023task,li2024tackling,wen2023task} primarily focus on the device-cloud collaborative inference.
This architecture, however, faces substantial challenges in terms of feasibility and scalability due to its reliance on a powerful centralized cloud server with high computational capability.
Moreover, prior works have generally employed the pipeline parallelism architectures, which are associated with inherent disadvantages such as pipeline bubbles \cite{brakel2024model}.
These bubbles occur when downstream devices are forced to remain idle while waiting for upstream computations to complete, leading to poor utilization of computational resources.
To address these limitations, distributed on-device LLM inference leveraging tensor parallelism has recently been proposed as a promising solution \cite{shoeybi2019megatron,dong2024towards,hansen2024communication}.
This approach divides large neural network tensors (e.g., weight matrices) of LLMs into smaller segments and distributes them across multiple edge devices.
It not only eliminates the reliance on a powerful central server but also enables concurrent processing of model segments across devices, significantly improving the utilization of computation and communication resources.
Nevertheless, a critical challenge in tensor parallelism is the frequent all-reduce operations required to aggregate intermediate layer outputs across devices.
These communication-intensive all-reduce steps can cause substantial latency in practical wireless networks and hinder real-time inference, necessitating efficient communication strategies to fully achieve the benefits of tensor parallelism.

In this paper, we propose a communication-efficient framework for distributed on-device LLM inference with tensor parallelism.
Specifically, we propose an over-the-air computation (AirComp) approach to facilitate fast all-reduce operations.
AirComp leverages the superposition property of wireless multiple-access channels, allowing simultaneous transmissions from multiple devices to be naturally summed at the receiver \cite{nazer2007computation,cui2005energy}.
This method reduces the communication latency and bandwidth requirement compared to traditional techniques that treat communication and computation separately.
Most recently, AirComp has gained popularity in various applications such as edge computing \cite{wang2022multi,frey2021over,cao2020optimized}, federated learning \cite{yang2020federated,zhu2024over,sery2021over}, and distributed sensing \cite{liu2023over,wang2024ultra,feres2023over}.
{\color{black}
Table \ref{tab:aircomp_inference} shows a thorough survey of recent state-of-the-art frameworks on distributed parallel computing and AirComp for both model training and inference tasks.
}

{\color{black}
	\renewcommand{\arraystretch}{1.4}
	\setlength{\tabcolsep}{2.7pt}
	
	\begin{table*}[t]
%		\vspace{5pt}
		\centering
		\small
		\color{black}
		\begin{tabular}{|c|c|c|c|c|c|c|}
			\toprule
			% 在表头中使用 \makecell 换行
			\textbf{\makecell[c]{Reference }} 
			& \textbf{\makecell[c]{Application\\ Scenario}} 
			& \textbf{\makecell[c]{Parallelism\\ Method}}
			& \textbf{\makecell[c]{Antenna\\ Configuration}}
			& \textbf{\makecell[c]{Optimization\\ Objective}}
			& \textbf{\makecell[c]{Large-Scale\\ LLM}}
			& \textbf{\makecell[c]{Device\\ Heterogeneity}}\\
			\midrule
			
			{K. Yang et al. (2020) \cite{yang2020federated}} 
			& Training 
			& {\makecell[c]{Data Parallelism}} 
			& Multi-Antenna
			& {\makecell[c]{Device Participation}}
			& \(\times\)
			& \(\times\) \\
			\midrule
			
			{ X. Fan et al. (2021) \cite{fanjoint}} 
			& Training 
			& Data Parallelism
			& Single-Antenna
			& Convergence Rate
			& $\times$
			& \(\checkmark\)\\
			\midrule
			
			{ T. Sery et al. (2021) \cite{sery2021over}} 
			& Training 
			& Data Parallelism
			& Single-Antenna
			& Communication Distortion
			& $\times$
			& \(\times\)\\
			\midrule
			
			{ Y. Liang et al. (2024) \cite{liangcommunicationaaa}} 
			& Training 
			& {\makecell[c]{Data Parallelism}}
			& Single-Antenna
			& {\makecell[c]{Training Latency and\\ Energy Consumption}}
			& $\times$
			& \(\checkmark\)\\
			\midrule
			
			{ H. Sun et al. (2024) \cite{sunfederated}} 
			& Training 
			& {\makecell[c]{Data and\\ Model Parallelism}}
			& Multi-Antenna
			& Convergence Rate
			& $\checkmark$
			& \(\checkmark\)\\
			\midrule
			
			{ Z. Zhuang et al. (2023) \cite{zhuangintegrated}} 
			& Inference
			& {\makecell[c]{Data and\\ Model Parallelism}}
			& Multi-Antenna
			& {\makecell[c]{Minimum Pair-Wise\\ Discriminant Gain}}
			& $\times$
			& \(\checkmark\)\\
			\midrule
			
			{ D. Wen et al. (2023) \cite{wen2023task} }
			& Inference
			& {\makecell[c]{Data Parallelism}}
			& Multi-Antenna
			& {\makecell[c]{Discriminant Gain}}
			& $\times$
			& \(\checkmark\)\\
			\midrule

			{ P. Yang et al. (2024) \cite{yangover} }
			& Inference
			& {\makecell[c]{Data and\\ Model Parallelism}}
			& Multi-Antenna
			& {\makecell[c]{Communication Distortion}}
			& $\times$
			& \(\times\)\\
			\midrule

			{This paper}
			& Inference
			& Tensor Parallelism
			& Multi-Antenna
			& Communication Distortion
			& \(\checkmark\)
			& \(\checkmark\) \\
			\bottomrule
		\end{tabular}
%		\vspace{2pt}
		\caption{\color{black}Overview of Over-the-Air Computation for Distributed Learning and Inference}\label{tab:aircomp_inference}
			\vspace{-4pt}
\end{table*}}

\begin{figure*}[t]
	%		\vspace{2.1pt}
	%	\hspace{-10pt}
	\renewcommand\figurename{\small Fig.}
	\centering \setlength{\baselineskip}{2pt}
	%	\hspace{-10pt}
	%	\hspace{-14.5pt}
	\includegraphics[width = 1\textwidth,trim=10 93 5 0,clip]{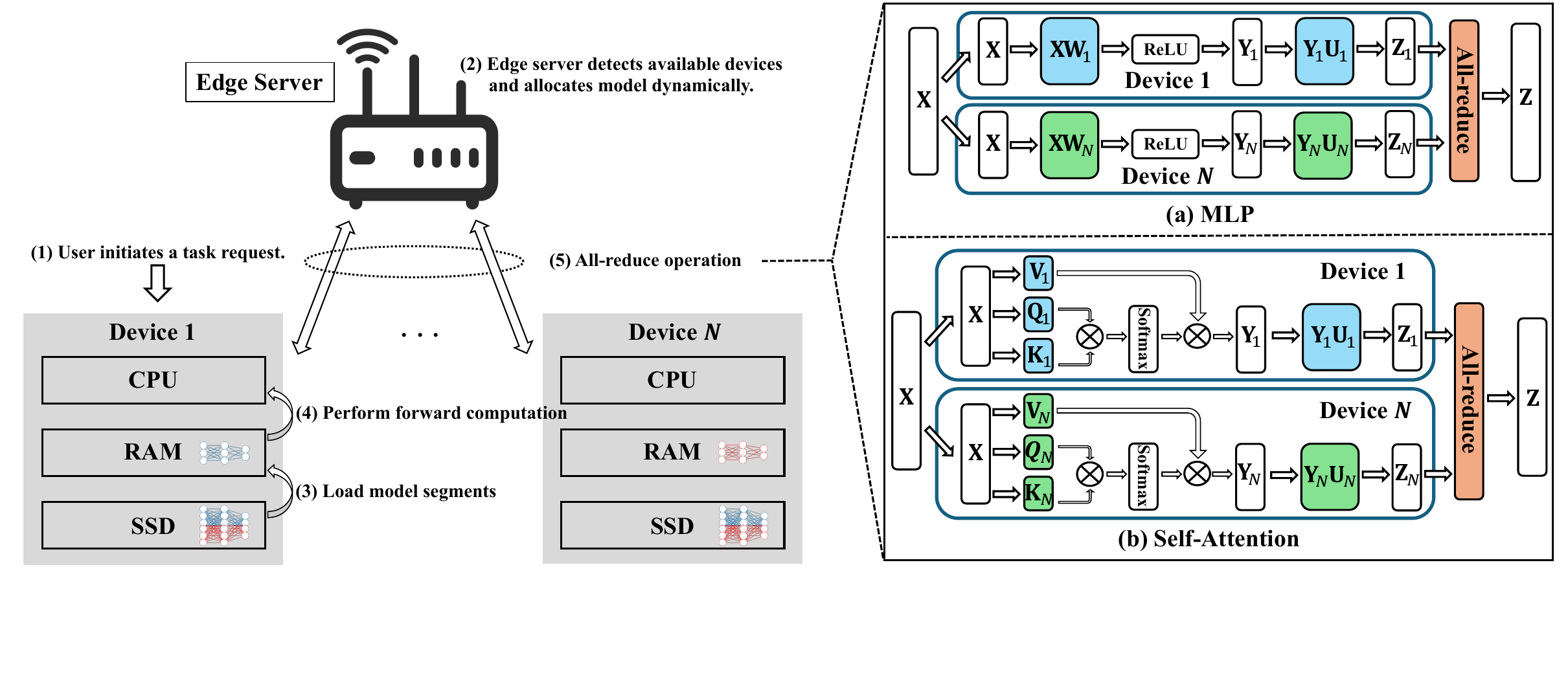}
	\vspace{2pt}
	\caption{An illustration of the distributed on-device LLM inference system, showing the system workflow and visualizing tensor parallelism for (a) MLP and (b) self-attention layers.}\label{illustration}
	\vspace{-5pt}
\end{figure*}

The performance of the proposed distributed LLM inference system, however, is heavily influenced by the communication efficiency, particularly given the limited energy resources of edge devices.
Thus, to improve the inference performance, we investigate a joint model assignment and transceiver optimization problem aimed at minimizing the average transmission mean-squared error (MSE).
The formulated joint optimization is crucial considering the heterogeneous computation capabilities of edge devices and varying wireless channel conditions.
Optimal model assignment ensures that each device processes a suitable portion of the model based on its computational capability (e.g., memory size and compute power), while transceiver optimization minimizes the communication distortions during the AirComp process.
To simplify the problem and gain key insights, we initially consider the scenario of single-antenna edge devices. We then extend the framework to a multi-antenna configuration, leveraging spatial multiplexing to further enhance communication efficiency and reduce inference latency.
Furthermore, the formulated joint model assignment and transceiver optimization problem is intractable due to its mixed-timescale, stochastic, and non-convex property. Specifically, the model assignment policy should be determined at the beginning of inference based on long-term statistical channel state information (CSI), while the transceiver design adapts dynamically to the CSI in each all-reduce step.
To address the mixed-timescale optimization problem, we develop an efficient two-stage algorithm by employing semidefinite relaxation (SDR) and stochastic successive convex approximation (SCA).
{\color{black} We note that although existing wireless optimization techniques (e.g., SDR and SCA algorithms) have been well studied, their tailored application to distributed LLM inference brings unique challenges and technical requirements. Specifically, our framework addresses unique challenges arising from large-scale distributed LLM inference, including the frequent aggregation of high-dimensional tensors, mixed-timescale optimization involving long-term model assignment and short-term transceiver adaptation, handling of heterogeneous device capabilities, multi-antenna AirComp beamforming designs, and stringent energy constraints.}

%\vspace{-5pt}
\subsection{Contributions}
The main contributions of this paper are summarized as follows.

	1) We propose a novel distributed on-device LLM inference framework by employing tensor parallelism and AirComp.
	While tensor parallelism effectively distributes computational workloads across edge devices, its frequent all-reduce operations incur significant communication overhead, which offsets the computational benefits and becomes a major bottleneck for inference performance.
	To address this challenge, we develop a communication-efficient AirComp all-reduce approach by exploiting the signal superposition property of wireless multiple-access channels.

	2) To utilize the heterogeneous computational capabilities of edge devices and mitigate communication distortions, we investigate a joint model assignment and transceiver optimization problem to minimize the average transmission MSE.
	The formulated mixed-timescale stochastic non-convex optimization problem is inherently intractable.
	Thus, we develop an efficient two-stage algorithm that decomposes the original problem into short-term transceiver optimization and long-term model assignment optimization subproblems.
	The resulting subproblems are further solved by employing SDR and stochastic SCA, respectively.
	The proposed algorithm requires no prior knowledge of channel statistics, and it converges almost surely to a stationary point of the original problem.

	3) We validate the effectiveness of the proposed framework through simulations with two state-of-the-art open-source LLMs and a real-world text dataset. Simulation results demonstrate that the proposed algorithm outperforms benchmark schemes across various network settings, achieving up to 5x inference speed acceleration and improving inference accuracy.

\subsection{Organization and Notations}
The rest of this paper is organized as follows. In Section \ref{sec_2}, we elaborate on the system model and present the problem formulation. In Section \ref{sec_3}, we develop a two-stage algorithm and prove its convergence. In Section \ref{sec_4}, we extend the algorithm for multi-antenna edge devices. Simulation results are presented in Section \ref{sec_5}, and we conclude the paper in Section \ref{sec_6}.

\emph{Notations:} Column vectors and matrices are denoted by boldface lowercase and boldface capital letters, respectively.
The symbol $\b{R}$ denotes the set of real numbers.
$\b{C}^{M\times N}$ represents the space of the $M \times N$ complex-valued matrices.
$(\cdot)^\s{T}$ and $(\cdot)^\s{H}$ stand for the transpose and the conjugate transpose of their arguments, respectively. $\textup{tr}(\bf{A})$ denote the trace of matrix $\bf{A}$.
$\b{E}[\cdot]$ denotes the expectation operation. $\nabla$ represents the gradient operator. $|\cdot|$ and $\|\cdot\|$ stand for the $\ell_1$ and $\ell_2$ norm of vectors.

\section{System Model and Problem Formulation}\label{sec_2}

In this section, we first elaborate on the proposed distributed on-device LLM inference system, followed by proposing the communication-efficient AirComp all-reduce approach.
To minimize the average transmission MSE,  we then formulate a joint model assignment and transceiver optimization problem.

\subsection{Distributed On-Device LLM Inference System}

To deploy LLMs on resource-limited edge devices, distributed on-device inference with tensor parallelism has been proposed.
This method involves partitioning large neural network tensors (e.g., weight matrices) of LLMs into smaller segments and distributing them across multiple edge devices for simultaneous processing.
The complete workflow of the proposed distributed on-device LLM inference system is illustrated in Fig. \ref{illustration}.
When a device initiates an inference request, the edge server dynamically identifies available local devices and partitions the model parameters. Then, each device loads its assigned model segment into memory and performs forward computation. After each layer of the LLM is computed, an all-reduce operation aggregates the intermediate layer outputs from all devices, ensuring synchronization and consistency across devices during inference.
{\color{black} In the proposed distributed inference framework, the device shares its input (typically token embeddings rather than raw data) with other participating devices.
	For scenarios demanding strict confidentiality, encryption schemes (e.g., homomorphic encryption) or secure enclaves can be adopted to mitigate privacy leakage.
%	Such approaches can be seamlessly combined with the proposed tensor-parallel design, albeit with additional computational overhead.
	Furthermore, we highlight two typical scenarios illustrating real-world, trusted environments particularly suitable for our distributed inference framework, as shown in the following.
	\begin{itemize}[leftmargin=10pt]
	\item \textbf{Organizational or HPC Clusters:} Large institutions (e.g., corporate data centers, national labs, or university HPC centers) often host massive LLMs that exceed the capacity of a single node. In these clusters, multiple servers within the same security domain can distribute model segments or layers among them, securely exchanging raw input data via internal networks. Since all compute nodes reside in the same trusted infrastructure (with well-defined access control, encryption, and compliance policies), they can fully leverage parallelization to reduce per-inference latency and alleviate memory bottlenecks, without risking data exposure to external environments.
	
	\item \textbf{Single-User or Local Edge Scenarios:} Individual users or small teams may possess multiple personal devices or localized edge servers (e.g., the home server or on-premises GPU node). These devices operate within a single-user network or closed local environment, allowing them to share raw inputs without breaching privacy. By splitting the LLM’s parameters or layers across these trusted devices, users can achieve faster response times and reduced memory load per device. These benefits are especially valuable for real-time applications (e.g., smart home assistants or AR/VR), where offloading data to external clouds may be undesirable or impractical.
	
	\end{itemize}
	}

%\vspace{-0.2pt}
\subsection{Tensor Parallelism}
LLMs are primarily built on the Transformer architecture, which typically consists of dozens of Transformer layers \cite{transformer}. Each Transformer layer includes a self-attention mechanism and a multi-layer perceptron (MLP).
To achieve efficient distributed inference, tensor parallelism partitions both the self-attention and MLP layers within each Transformer block into smaller tensor segments, as shown in Fig. \ref{illustration}.
{\color{black} We note that both pipeline parallelism and tensor parallelism are two prevalent model partitioning strategies widely adopted in distributed inference frameworks. While pipeline parallelism partitions the model across layers, tensor parallelism partitions computations within each layer across multiple devices. Tensor parallelism is particularly attractive for on-device inference due to its inherent advantages in significantly reducing idle times (pipeline bubbles), achieving finer-grained memory allocation, and, when combined with AirComp-based aggregation, greatly minimizing communication overhead. These properties highlight the practical benefits and superior suitability of tensor parallelism for resource-constrained and latency-sensitive inference scenarios considered in this work.}

\subsubsection{Tensor Parallelism for MLP Layer}

\begin{figure}[t]
	\renewcommand\figurename{Fig.}
	\centering \vspace*{1pt} \setlength{\baselineskip}{10pt}
	\fbox{\includegraphics[width = 0.47\textwidth,trim=0 0 60 0,clip]{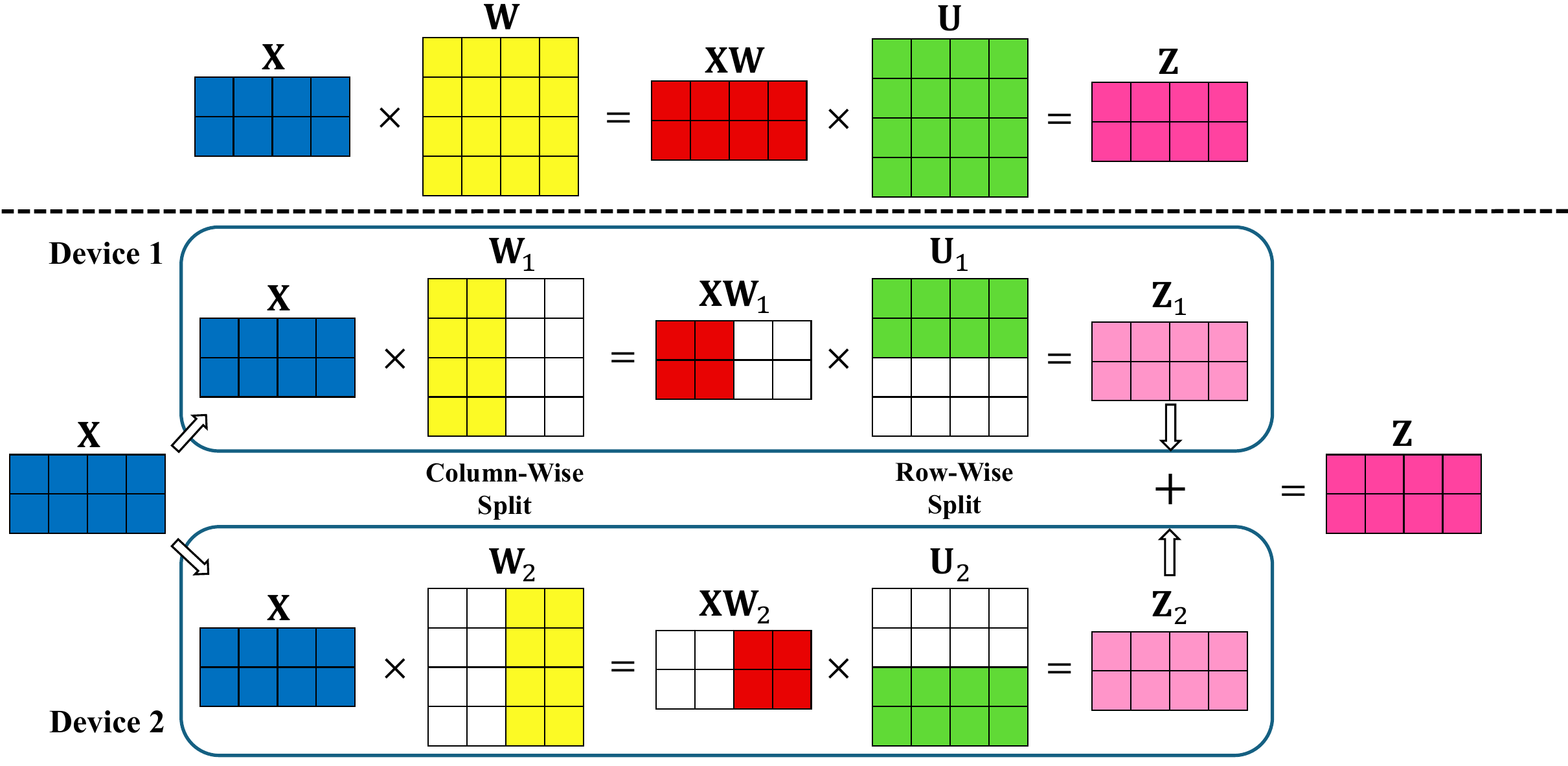}}
	%	\vspace*{-2pt}
	\caption{\color{black}Illustration of MLP matrix multiplication for conventional unpartitioned approach and tensor parallelism with two devices.}\label{fig_mlp}
	\vspace{-10pt}
\end{figure}

For a typical 2-layer MLP within the Transformer block, the forward computation involves two main linear transformations, separated by a non-linear activation function (e.g., ReLU or GeLU). 
{\color{black}
We formulate the computation of the MLP layer by taking the ReLU activation as an example.
Mathematically, it is expressed as follows,
\begin{equation}
%	\vspace{2.1pt}
	\begin{aligned}
		\bf{Z} = \max(\bf{0}, \bf{XW})\bf{U},
	\end{aligned}    
\end{equation}
where $\bf{X}$ is the input to the MLP layer, $\bf{Z}$ is the output, and $\bf{W}$ and $\bf{U}$ are the weight matrices, respectively.
Our framework can be readily generalized to other activation functions, such as GeLU function: $
	\mathbf{Z} = \text{GeLU}(\mathbf{XW})\mathbf{U},~ \text{where}~\text{GeLU}(x)=x\Phi(x),$
with $\Phi(x)$ representing the cumulative distribution function of the standard Gaussian distribution.}
The traditional centralized inference approach loads the entire weight matrices $\bf{W}$ and $\bf{U}$ into memory and performs full matrix multiplications on a single device, which is usually impractical for resource-limited edge devices.
To overcome this challenge, tensor parallelism distributes the weight matrices $\bf{W}$ and $\bf{U}$ across $N$ devices.
{\color{black} The weight matrices $\mathbf{W}$ and $\mathbf{U}$ have dimensions $\displaystyle d \times d_{\text{hidden}}$ and $\displaystyle d_{\text{hidden}} \times d$, respectively.
As shown in Fig. \ref{fig_mlp}, the weight matrix $\mathbf{W}$ is partitioned column-wise into multiple slices as
\begin{equation}
	%\vspace{2.1pt}
	\begin{aligned}
		&\bf{W} \\
		&= [\bf{W}_1 \in \b{R}^{d \times d^1_{\text{hidden}}}, \bf{W}_2\in \b{R}^{d \times d^2_{\text{hidden}}}, \ldots, \bf{W}_N\in \b{R}^{d \times d^N_{\text{hidden}}}],
	\end{aligned}    
\end{equation}
where $\bf{W}_n$ represents the portion of the weight matrix $\bf{W}$ assigned to device $n$ and $d_{\text{hidden}} = \sum_{n=1}^{N}d^n_{\text{hidden}}$. Similarly, the weight matrix $\mathbf{U}$ is partitioned row-wise as
\begin{equation}
	%\vspace{2.1pt}
	\begin{aligned}
		\bf{U} = \begin{bmatrix}
			\bf{U}_1 \in \b{R}^{ d^1_{\text{hidden}}  \times d}\\
			\bf{U}_2 \in \b{R}^{ d^2_{\text{hidden}}  \times d} \\
			\ldots\\
			\bf{U}_N \in \b{R}^{ d^N_{\text{hidden}}  \times d}
		\end{bmatrix},
	\end{aligned}    
\end{equation}
where $\bf{U}_n$ represents the portion of the weight matrix $\bf{U}$ assigned to device $n$.}
% as illustrated below,
%\vspace{2.1pt}
%\begin{equation}
%%\vspace{2.1pt}
%	\begin{aligned}
%		\bf{W} &= [\bf{W}_1, \ldots, \bf{W}_N],\\
%		\vspace{6pt}
%		\bf{U} &= [\bf{U}_1^{\s{T}}, \ldots, \bf{U}_N^{\s{T}}]^{\s{T}},
%	\end{aligned}    
%\end{equation}
%where $\bf{W}_n$ and $\bf{U}_n$ represent the portions of the weight matrices assigned to device $n$.
Then, each device $n$ can perform the forward computation on its respective model segment as follows,
%\vspace{1.1pt}
\begin{equation}
%\vspace{1.1pt}
	\begin{aligned}
		\bf{Z}_n = \mathrm{max}(0, \bf{X} \bf{W}_n) \bf{U}_n,
	\end{aligned}    
\end{equation}
where $\bf{Z}_n$ is the partial output produced by device $n$. 
Once all devices obtain their local outputs $\bf{Z}_n$, an all-reduce operation is performed to aggregate the partial outputs from all devices as follows,
\vspace{2.1pt}
\begin{equation}\label{aggre_rule}
\vspace{2.1pt}
	\begin{aligned}
		\bf{Z} = \sum_{n=1}^N \bf{Z}_n.
	\end{aligned}    
\end{equation}
{\color{black}
The validity of this aggregation can be explained by considering how the model parameters are partitioned across the devices. Specifically, concatenating the column slices $\mathbf{W}_n$ reproduces $\mathbf{W}$, and stacking the row slices $\mathbf{U}_n$ recovers $\mathbf{U}$. Consequently, the original unpartitioned MLP output $\bf{Z}$ can be expressed as
\vspace{4pt}
\begin{equation}
	\begin{aligned}
		\bf{Z} &= \max(\bf{0}, \bf{XW})\bf{U} \\
		&= \max(\bf{0}, \bf{X[\bf{W}_1 , \bf{W}_2, \ldots, \bf{W}_N]})\begin{bmatrix}
			\bf{U}_1 \\
			\bf{U}_2 \\
			\ldots\\
			\bf{U}_N 
		\end{bmatrix} \\
		&\aeq \sum_{n=1}^N  \mathrm{max}(0, \bf{X} \bf{W}_n) \bf{U}_n \\
		&= \sum_{n=1}^N \bf{Z}_n,
	\end{aligned}    
\end{equation}
where (a) follows from the element-wise property of activation functions (e.g., ReLU, GeLU). Therefore, aggregating partial results $\mathbf{Z}_n$ reconstructs the original unpartitioned output $\mathbf{Z}$ (i.e., Eq. \eqref{aggre_rule} holds).}
After aggregation, the final output $\bf{Z}$ of the MLP layer is broadcasted to all devices, ensuring synchronization and consistency across devices for the subsequent layer's computation.

\subsubsection{Tensor Parallelism for Self-Attention Layer}
For the self-attention layer, tensor parallelism similarly partitions its query ($\bf{Q}$), key ($\bf{K}$), value ($\bf{V}$), and transformation ($\bf{U}$) matrices across edge devices.
In the traditional centralized computation of the self-attention layer, the output $\bf{Z}$ can be derived as follows,
\vspace{2.1pt}
\begin{equation}
	\vspace{2.1pt}
	\begin{aligned}
		\bf{Z} = \mathrm{softmax} \left( \frac{\bf{X}\bf{Q} (\bf{X}\bf{K})^{\s{T}}}{\sqrt{d_k}} \right)\bf{V}\bf{U},
	\end{aligned}    
\end{equation}
where $\bf{X}$ denotes the input, and $d_k$ denotes the dimension of the key vectors.
In tensor parallelism, the memory-intensive weight matrices are splited and distributed across $N$ edge devices as follows, 
\begin{equation}
%	\vspace{2.1pt}
	\begin{aligned}
		\bf{Q} &= [\bf{Q}_1, \ldots, \bf{Q}_N],\\
		\bf{K} &= [\bf{K}_1, \ldots, \bf{K}_N],\\
		\bf{V} &= [\bf{V}_1, \ldots, \bf{V}_N],\\
		\bf{U} &= [\bf{U}_1^{\s{T}}, \ldots, \bf{U}_N^{\s{T}}]^{\s{T}}.
	\end{aligned}    
\end{equation}
Then, each device $n$ performs local computation on its corresponding portion of the query, key, value, and transformation matrices as follows,
\vspace{2.1pt}
\begin{equation}
	\vspace{2.1pt}
	\begin{aligned}
		\bf{Z}_n = \mathrm{softmax} \left( \frac{\bf{X}\bf{Q}_n (\bf{X}\bf{K}_n)^{\s{T}}}{\sqrt{d_k}} \right)\bf{V}_n\bf{U}_n.
	\end{aligned}    
\end{equation}
Once all devices obtain their local outputs $\bf{Z}_n$, a similar all-reduce operation is required to gather and combine the partial outputs from devices as shown in \eqref{aggre_rule}.

%\vspace{-1spt}
\subsection{Over-the-Air All-Reduce}
%\vspace{-1.1pt}
Employing tensor parallelism for distributed LLM inference requires frequent all-reduce operations, which cause significant communication overhead in practical wireless networks.
To address this issue, we propose a communication-efficient AirComp all-reduce approach. The AirComp aggregates distributed data efficiently by leveraging the signal superposition property of wireless multiple-access channels, allowing simultaneous transmissions to compute nomographic functions (e.g., arithmetic mean) \cite{goldenbaum2013harnessing}. In the proposed distributed LLM inference system, the aggregation of intermediate layer outputs in the all-reduce step aligns with this operation, making the AirComp suitable to mitigate communication overhead. {\color{black}Note that edge devices performing AirComp must achieve symbol-level synchronization to ensure their transmitted signals arrive concurrently at the receiver, minimizing aggregation errors due to timing offsets. In our framework, synchronization among edge devices can be practically realized through the well-established timing advance (TA) mechanism. Specifically, the edge server estimates each device’s timing offset and instructs each device to adjust its signal transmission timing via dedicated TA commands. By aligning transmissions precisely, edge devices can ensure simultaneous arrival and accurate signal aggregation at the receiver.}

We consider a wireless network consisting of an edge server with $N_r$ antennas and $N$ single-antenna edge devices.
We further extend the proposed framework to a more general scenario invloving multi-antenna edge devices in Section \ref{sec_4}.
The uplink channels from edge devices to the server are block-fading, where channel statistics remain constant throughout the inference process, with channel states varying independently across different time intervals.
Let $z_n$ denote the per-round transmitted entry of device $n$'s intermediate layer output $\bf{Z}_n$, which has a complete dimensionality of $L_0$.
To reduce transmission power, the transmitted symbols are normalized to have zero mean and unit variance, i.e., $\mathbb{E}[\|z_n\|^2]=1$, where the normalization factor is uniform for all devices and can be inverted at the server.
Given synchronized symbol boundaries, all devices transmit their intermediate layer outputs simultaneously. 
To mitigate the distortion of received signals caused by channel noise, aggregation beamforming is adopted.
Let $\mathbf{a} \in \mathbb{C}^{N_r \times 1}$ denote the aggregation beamforming vector at the edge server.
%{\color{black}AirComp leverages the coherent superposition of analog signals transmitted simultaneously from multiple devices. Therefore, devices must achieve symbol-level synchronization to ensure their transmitted signals arrive concurrently at the receiver, minimizing aggregation errors due to timing offsets.}
After the AirComp, the received signal at the server is given by,
\begin{equation}\label{rec_scalar}
	\begin{aligned}
		&\hat{z}  = \bf{a}^{\s{H}}\sum_{n=1}^{N}\mathbf{h}_n b_n z_n + \bf{a}^{\s{H}} \bf{n},
	\end{aligned}    
\end{equation}
where $\bf{h}_n\in \b{C}^{N_r\times 1}$ denotes the uplink channel from device $n$ to the server, $b_n$ is the transmit power of device $n$, and $\bf{n}\sim \mathcal{CN}\left(0, \sigma^{2} \mathbf{I}\right)$ denotes the additive white Gaussian noise vector with $\sigma^{2}$ being the noise power.
{\color{black} In the single-antenna setting, each device employs only a scalar transmit-power coefficient $b_n$ (instead of a beamforming vector) to scale its transmitted scalar entry $z_n$.}
The distortion of $\hat{z}$ with respect to the desired target summation $z  = \sum_{n=1}^{N} z_n$ is measured by the MSE, which is defined as
\begin{equation}\label{mse_scalar}
	\begin{aligned}
		\textup{MSE}(\hat{z},z) = \mathbb{E}\left[\|\hat{z}-z\|^2\right].
	\end{aligned}    
\end{equation}
The MSE serves as a metric to evaluate the performance of the AirComp all-reduce operations.
As shown in the simulations later, the inference accuracy of the distributed on-device LLM inference system is greatly influenced by the transmission error during the AirComp phase.
By substituting \eqref{rec_scalar} into \eqref{mse_scalar}, the MSE can be explicitly represented as a function of aggregation beamforming vector $\mathbf{a}$ and transmitter scalars $\left\lbrace b_n  \right\rbrace_{n=1}^N$ as follows,
%\vspace{1.1pt}
\begin{equation}
	\vspace{1.1pt}
	\begin{aligned}
		&\textup{MSE}(\mathbf{a},\left\lbrace b_n  \right\rbrace ) = \sum_{n=1}^{N}   \left\| \mathbf{a}^{\s{H}}  \mathbf{h}_n b_n - 1 \right\|^2 + \sigma^2  \mathbf{a}^{\s{H}} \mathbf{a}.   \\
	\end{aligned}    
\end{equation}

Edge devices involved in inference tasks typically have limited energy supply. Thus, we assume that for each device $n$, the energy consumption for both the forward computation of each LLM layer and the transmission of the intermediate output cannot exceed the maximum power budget $P_{n}^{\textup{max}}$.
To model the computation energy consumption, we first introduce a model assignment vector $\bf{m}=[m_1,\ldots,m_N]$ with its entry $m_n \in [0,1]$ representing the proportion of model allocated to device $n$.
Consequently, the computation energy consumption for device $n$ is given by $e_n m_n s^{\textup{tot}}$, where $e_n$ denotes the device-specific energy coefficient that reflects the energy cost associated with accessing and processing each weight during computation, and $s^{\textup{tot}}$ is the number of parameters (weights) for each layer. The communication energy consumption of device $n$ can be derived as $L_0 \|  b_n \|^2$. Accordingly, the power constraint is given by
\vspace{1.1pt}
\begin{equation}
%	\vspace{2.1pt}
	\begin{aligned}
		e_n m_n s^{\textup{tot}} + L_0 \|  b_n \|^2\leq P_{n}^{\textup{max}} , \forall n.
	\end{aligned}    
\end{equation}

\subsection{Problem Formulation}

In the proposed distributed LLM inference system, the overall performance is determined by the model assignment policy $\bf{m}$ and the transceiver design $\mathbf{a}$, $\left\lbrace b_n \right\rbrace$.
Optimal model assignment ensures that each device processes a suitable portion of the model based on its computational capability (e.g., memory size and compute power). Meanwhile, efficient transceiver optimization can reduce signal misalignment error and suppress channel noise, thereby improving inference accuracy.
Thus, to improve inference performance, we formulate a joint model assignment and transceiver optimization problem that aims to minimize the average MSE, subject to the per-device power constraints.
Importantly, the transceiver design can adapt dynamically to instantaneous CSI.
In contrast, adapting the model assignment policy to instantaneous CSI in a real-time manner is impractical due to the significant latency caused by loading different model segments. Thus, model assignment should be finished before inference based on the long-term channel statistics.

The resulting problem is therefore formulated as a mixed-timescale joint optimization of the short-term transceiver variables $\mathbf{a}$, $\left\lbrace b_n \right\rbrace$ and the long-term model assignment policy $\mathbf{m}$ as follows,
\begin{equation}
	\begin{aligned}
		\ca{P}_1:~\min_{\bf{m}} &~~ \mathbb{E}_{\mathbf{h}}\left[ \min_{\mathbf{a},\left\lbrace b_n  \right\rbrace} \textup{MSE}(\mathbf{a},\left\lbrace b_n  \right\rbrace )  \right]  \\
		\textup{s.t.}& ~~ e_n m_n s^{\textup{tot}} + L_0 \|  b_n \|^2\leq P_{n}^{\textup{max}} , \forall n,\\
		&~ ~\sum_{n=1}^{N}m_n=1,\\
		& ~~0 \leq m_n \leq 1, \forall n,
	\end{aligned}    
\end{equation}
where the expectation $\mathbb{E}_{\mathbf{h}}\left[ \cdot \right] $ is taken over all random channel realizations $\mathbf{h} = \left\lbrace \bf{h}_n \right\rbrace_{n=1}^N $. 
However, the problem $\ca{P}_1$ is challenging to be solved due to the following three reasons.
\begin{itemize}
	\item Non-convexity: The objective function is inherently non-convex due to the coupling between the receiver aggregation beamformer $\mathbf{a}$ and the transmitter scalars $\left\lbrace b_n \right\rbrace$.
	\item Expectation over Random Channels: The objective involves an expectation over random CSI, which requires prior knowledge of channel statistics.
	\item Interdependence of Timescales: The per-device power constraints link the short-term transceiver variables with the long-term model assignment policy, leading to a complex interplay between the two timescales.
\end{itemize}
To address these challenges, we develop a two-stage algorithm that separately solves the short-term transceiver optimization and the long-term model assignment optimization in the following section.

\section{Algorithm Development}\label{sec_3}

In this section, we develop an efficient two-stage algorithm to solve the joint model assignment and transceiver optimization problem $\ca{P}_1$.
Then, we show that the proposed algorithm can converge to a stationary point of the original problem $\ca{P}_1$.

\subsection{Problem Decomposition}

We start by decomposing problem $\ca{P}_1$ into a family of short-term transceiver optimization problems and a long-term model assignment optimization problem as follows.

\subsubsection{Short-term transceiver optimization for given model assignment policy $\bf{m}$ and channel condition $\bf{h}$}
%\vspace{2.1pt}
%\vspace{-1.5pt}
\begin{equation}
	%	\vspace{-1.5pt}
	\begin{aligned}
		\ca{P}_{s}: \min_{\mathbf{a},\left\lbrace b_n  \right\rbrace}  &~ \textup{MSE}(\mathbf{a},\left\lbrace b_n  \right\rbrace )   \\
		\textup{s.t.}~~& ~ e_n m_n s^{\textup{tot}} + L_0 \|  b_n \|^2\leq P_{n}^{\textup{max}} , \forall n.
	\end{aligned}    
\end{equation}

\subsubsection{Long-term model assignment optimization based on the optimal solution $\mathbf{a}^{*}(\bf{m}),\left\lbrace b_n^{*}(\bf{m})  \right\rbrace$ to problem $\ca{P}_{s}$}

\begin{equation}
%\vspace{-1.5pt}
%\vspace{2.1pt}
\begin{aligned}
	%		\vspace{-1.5pt}
	\ca{P}_l:\min_{\bf{m}} &~~ \mathbb{E}_{\mathbf{h}}\left[ \textup{MSE}(\mathbf{a}^{*}(\bf{m}),\left\lbrace b_n^{*}(\bf{m})  \right\rbrace )    \right]  \\
	\textup{s.t.}& ~~ e_n m_n s^{\textup{tot}} + L_0 \|  b_n^{*}(\bf{m}) \|^2\leq P_{n}^{\textup{max}} , \forall n, \\
	&~ ~\sum_{n=1}^{N}m_n=1,\\
	& ~~0 \leq m_n \leq 1 , \forall n.
\end{aligned}
\end{equation}
The short-term transceiver optimization problem $\ca{P}_s$ remains non-convex, and we address it using the SDR technique. The long-term model assignment optimization problem $\ca{P}_l$ is similarly challenging, as the optimal transceiver variables $\mathbf{a}^{*}(\bf{m}),\left\lbrace b_n^{*}(\bf{m})  \right\rbrace$ cannot be derived in closed form. Additionally, the distribution of CSI is difficult to obtain in practical wireless systems. To address these challenges, we propose a stochastic SCA algorithm that operates without requiring prior knowledge of channel statistics. In the following subsections, we provide a detailed implementation of the proposed algorithms.

\subsection{Short-Term Transceiver Optimization for $\ca{P}_s$}

The short-term problem $\ca{P}_{s}$ is challenging to be solved due to the inherent non-convexity caused by the coupling between receiver aggregation beamformer $\mathbf{a}$ and the transmitter scalers $\left\lbrace b_n \right\rbrace_{n=1}^N$. Thus, we first simplify problem $\ca{P}_{s}$ by demonstrating that the channel inversion precoding is optimal conditioned on the aggregation beamformer.

\begin{lem}\label{lem_opt_precoder_s}
	For a given aggregation beamformer $\bf{a}$, the transmission MSE is minimized by using the zero-forcing precoders $b_n^* =\frac{1}{\bf{a}^{\s{H}}\bf{h}_n}, \forall n$.
\end{lem}
\begin{proof}
	{Lemma \ref{lem_opt_precoder_s} can be proved by following the same steps as in \cite[Appendix A]{li2019wirelessly} and we omit it for brevity.}
\end{proof}

Let $\bf{g}$ represent the normalized aggregation beamformer that satisfies $\bf{g}^{\s{H}}\bf{g}=1$, and consequently $\bf{a}=\sqrt{\alpha} \bf{g}$ where $\alpha$ is optimized to satisfy the power constraints of edge devices.
By applying Lemma \ref{lem_opt_precoder_s}, problem $\ca{P}_{s}$ can be reformulated as follows,
%\vspace{1.1pt}
\begin{equation}\label{p_s_1_s}
	\begin{aligned}
		\min_{\alpha, \bf{g}} &~~ \alpha\\
		\textup{s.t.} & ~~ e_n m_n s^{\textup{tot}} + \frac{L_0}{\alpha \| \bf{g}^{\s{H}} \bf{h}_n \|^2}  \leq P_{n}^{\textup{max}}, \forall n,\\
		& ~~\bf{g}^{\s{H}}\bf{g}= 1.
	\end{aligned}  
\end{equation}
Then, by employing the equation $\| \bf{g}^{\s{H}} \bf{h}_n \|^2 =  \textup{tr}\left( \bf{h}_n \bf{h}_n^{\s{H}}  \bf{g} \bf{g}^{\s{H}}\right) $, an equivalent formulation of problem \eqref{p_s_1_s} is obtained as follows,
%\vspace{1.1pt}
\begin{equation}\label{p_s_2_s}
%	\vspace{1.1pt}
	%	\vspace{-1.5pt}
	\begin{aligned}
		\min_{\alpha, \bf{g}} &~~ \alpha\\
		\textup{s.t.} & ~~ e_n m_n s^{\textup{tot}} + \frac{L_0}{\alpha \textup{tr}\left( \bf{h}_n \bf{h}_n^{\s{H}}  \bf{g} \bf{g}^{\s{H}}\right)}  \leq P_{n}^{\textup{max}}, \forall n,\\
		& ~~\bf{g}^{\s{H}}\bf{g}= 1.
	\end{aligned}  
\end{equation}
The problem \eqref{p_s_2_s} remains intractable due to the non-convex norm constraint on $\bf{g}$. To address this issue, we apply the SDR approach that relaxes the non-convex norm constraint by employing its convex hull.

\begin{lem}\label{lem_cvx_hull}
	(Convex Hull Relaxation \cite{overton1992sum})
	Suppose the set $\Omega_1 = \lbrace \bf{Y}:\bf{Y}= \bf{X} \bf{X}^{\s{H}}, \bf{X}^{\s{H}}\bf{X} = \bf{I}_d  \rbrace $ and set $\Omega_2 = \left\lbrace \bf{Y}:\textup{tr} ( \bf{Y} ) = d, 0 \preceq \bf{Y} \preceq \bf{I}  \right\rbrace $, where $\bf{Y}$ is of the size $m$ by $m$ while $\bf{X}$ is of the size $m$ by $d$. The condition of the set $\Omega_2$ indicates that both $\bf{Y}$ and $\bf{I} - \bf{Y}$ are positive semi-definite. Then, $\Omega_2$ is the convex hull of $\Omega_1$, and $\Omega_1$ is the set of extreme points of $\Omega_2$.
\end{lem}

By applying Lemma \ref{lem_cvx_hull}, we can replace the non-convex norm constraint by its convex hull and reformulate a relaxed version of problem \eqref{p_s_2_s} as follows,
\begin{equation}\label{p_s_3_s}
%	\vspace{2.1pt}
	%	\vspace{-1.5pt}
	\begin{aligned}
		\min_{\alpha, \hat{\bf{G}}} &~~ \alpha\\
		\textup{s.t.} & ~~ e_n m_n s^{\textup{tot}} + \frac{L_0}{\alpha \textup{tr}\left( \bf{h}_n \bf{h}_n^{\s{H}}  \bf{g} \bf{g}^{\s{H}}\right)}  \leq P_{n}^{\textup{max}}, \forall n,\\
		& ~~\textup{tr}( {\hat{\bf G}} ) = 1 ,\\
		& ~~0 \preceq \bf{{\hat{\bf G}}} \preceq \bf{I} ,
	\end{aligned}  
\end{equation}
where ${\hat{\bf G}} = \bf{g}\bf{g}^{\s{H}}$. The problem \eqref{p_s_3_s} can be proved to be convex, and the globally optimal solution $\hat{\bf{G}}^*$ can be obtained by using a convex solver (e.g., the CVX toolbox in MATLAB \cite{grant2014cvx}).

We note that the optimal solution $\hat{\bf{G}}^*$ has a high probability to satisfy the rank-one constraint \cite{li2019wirelessly}. If a rank-one solution $\hat{\bf{G}}^*$ is obtained, the optimal solution $\bf{g}^{*}$ of the original problem \eqref{p_s_2_s} can be immediately achieved by extracting the dominant eigenvector of $\hat{\bf{G}}^*$ as $\bf{g}^{*}=[\bf{V}_{\hat{\bf{G}}^*}]_{:,1}$. Otherwise, if the rank of $\hat{\bf{G}}^*$ is larger than 1, we apply the Gaussian randomization algorithm \cite{luo2010semidefinite} to map the solution to a feasible, near-optimal solution for the original non-convex problem.

\subsection{Long-Term Model Assignment Optimization for $\ca{P}_l$}
%\vspace{-1pt}

In this subsection, we propose a stochastic SCA algorithm to solve the long-term model assignment problem $\ca{P}_l$. The proposed algorithm requires no prior knowledge of channel statistics.
For clearer algorithmic description, we first reformulate the long-term problem $\ca{P}_l$ into an equivalent form as follows,
%\vspace{2.1pt}
%\vspace{-2pt}
\begin{equation}\label{eee}
%	\vspace{2.1pt}
%	\vspace{-3pt}
	\begin{aligned}		
		\min_{\bf{m}} &~~ f_0(\bf{m})=\mathbb{E}_{\mathbf{H}}\left[ \textup{MSE}(\mathbf{a}^{*}(\bf{m}),\left\lbrace b_n^{*}(\bf{m})  \right\rbrace )    \right]  \\
		%		\vspace{-10pt}
		\textup{s.t.}& ~~ f_1(\bf{m})= s^{\textup{tot}} \textup{diag}(\bf{e} \bf{m}^{\s{T}})  + L_0 \bf{e}_c\left(\bf{m} \right) \leq \bf{p}^{\textup{max}},\\
		&~ ~\sum_{n=1}^{N}m_n=1,\\
		& ~~0 \leq m_n \leq 1 , \forall n,
	\end{aligned}    
\end{equation}
where $\bf{e}_c\!\left(\bf{m} \right) \!=\! [\|  b_1^{*}(\bf{m}) \|^2, \ldots, \|  b_N^{*}(\bf{m}) \|^2]^{\s{T}}\!,$
$\bf{p}^{\textup{max}}=[{P}_1^{\textup{max}}, \ldots, {P}_N^{\textup{max}}]^{\s{T}}$, and $\bf{e} = [e_1,\ldots,e_N]^{\s{T}}$.
The proposed stochastic SCA algorithm iteratively performs the following two steps: First, quadratic surrogate functions $\hat{f}_0(\bf{m})$, $\hat{f}_1(\bf{m})$ are constructed to approximate the non-convex components of the original objective and constraint functions $f_0(\bf{m})$, $f_1(\bf{m})$, respectively.
Then, the resulting convex quadratic approximation problem is solved, and the long-term model assignment policy is updated based on the solution. The details of these two steps are illustrated as follows.

\subsubsection{Step 1}

In each iteration $\tau$, the edge server first generates a channel sample $\bf{h}^{\tau}$, and then calculates the short-term transceiver variables $\mathbf{a}^{*}(\bf{m}^\tau)$ and $\left\lbrace b_n^{*}(\bf{m}^\tau)  \right\rbrace_{n=1}^N$ by solving the short-term problem $\ca{P}_s$.
Then, the recursive convex approximation of the original objective function $f_0(\bf{m})$ can be derived as{\color{black}\cite{liu2019stochastic}}
\begin{equation}\label{surrogate_obj_function}
	\begin{aligned}
		\hat{f}_0^{\tau}(\bf{m}) = \bar{f}_0(\bf{m}^{\tau}) + (\bf{u}_0^{\tau})^{\s{T}}\left( \bf{m} - \bf{m}^{\tau} \right)  + \eta_0 \left\|  \bf{m} - \bf{m}^{\tau} \right\|^2,
	\end{aligned}    
\end{equation}
where $\eta_0$ is a constant that ensures convexity, $\bar{f}_0(\bf{m}^{\tau})=\textup{MSE}(\mathbf{a}^{*}(\bf{m}^{\tau}), \left\lbrace b_n^{*}(\bf{m}^{\tau}) \right\rbrace)$ denotes the sample-wise approximation of the average MSE and is computed by using the specific channel realization $\bf{h}^{\tau}$. 
Furthermore, $\bf{u}_0^{\tau}$ is an approximation of the gradient $\nabla f_0(\bf{m}^{\tau})$, which is updated recursively as
\vspace{2.1pt}
\begin{equation}
	\vspace{2.1pt}
	\begin{aligned}
		\bf{u}_0^{\tau} = (1 - \rho^\tau) \bf{u}_0^{\tau-1} + \rho^\tau \nabla_{\bf{m}} \bar{f}_0(\bf{m}; \mathbf{a}^{*}(\bf{m}^\tau),\left\lbrace b_n^{*}(\bf{m}^\tau)  \right\rbrace),
	\end{aligned}
\end{equation}
and $\bf{u}_0^{-1} = \bf{0}${\color{black}\cite{liu2019stochastic}}. The algorithm parameter $ \rho^\tau$ is decreasing in $\tau$, satisfying $\lim_{\tau \rightarrow\infty}\rho^{\tau}=0$, $\sum_{\tau=0}^{\infty}\rho^{\tau}=\infty$, $\sum_{\tau=0}^{\infty}(\rho^{\tau})^2<\infty$, and $\sum_{\tau=0}^{\infty} \rho^{\tau}\tau^{-1/2}<\infty$.
Similarly, the recursive convex approximation of the power constraint function $f_1(\bf{m})$ is given by
\vspace{2.1pt}
\begin{equation}\label{surrogate_constraint_function}
	\vspace{2.1pt}
	\begin{aligned}
		\hat{f}_1^{\tau}(\bf{m}) = f_1(\bf{m}^{\tau}) + (\bf{u}_1^{\tau})^{\s{T}}\left( \bf{m} - \bf{m}^{\tau} \right)  + \eta_1 \left\|  \bf{m} - \bf{m}^{\tau} \right\|^2,
	\end{aligned}    
\end{equation}
where $\eta_1 >0$ is a constant, and $\bf{u}_1^{\tau}$ is updated recursively as follows{\color{black}\cite{liu2019stochastic}},
\vspace{2.1pt}
\begin{equation}
	\vspace{2.1pt}
	\begin{aligned}
		\bf{u}_1^{\tau} = (1 - \rho^\tau) \bf{u}_1^{\tau-1} + \rho^\tau \nabla_{\bf{m}} f_1(\bf{m}; \mathbf{a}^{*}(\bf{m}^\tau),\left\lbrace b_n^{*}(\bf{m}^\tau)  \right\rbrace).
	\end{aligned}
\end{equation}
{\color{black}It is noted that the surrogate functions $\hat{f}_0^\tau(\mathbf{m})$ and $\hat{f}_1^\tau(\mathbf{m})$ are quadratic approximations of the original nonconvex objective ${f}_0(\mathbf{m})$ and constraint ${f}_1(\mathbf{m})$ around the current iterate $\mathbf{m}^{\tau}$. Specifically, at iteration $\tau$, each surrogate function is constructed using first-order Taylor expansions of the corresponding function $f_i(\mathbf{m})$ (for $i=0,1$), along with an additional quadratic regularization term controlled by convexity constants $\eta_0$ and $\eta_1$. The convexity constants $\eta_0$ and $\eta_1$ serve to ensure strong convexity and numerical stability of the surrogate functions. Specifically, larger values of $\eta_0$ and $\eta_1$ enhance numerical stability but may slow convergence, whereas smaller values permit larger update steps but require careful tuning to prevent instability. In practice, setting these constants within the range $10^{-2}\sim10^{-1}$ (e.g., around 0.05) achieves a favorable balance between stability and convergence speed.}

\subsubsection{Step 2}

After obtaining the convex approximations of the objective and constraint functions, we formulate a convex approximation of the original problem \eqref{eee} to solve the optimal $\hat{\bf{m}}^{\tau}$ as follows,
\begin{equation}\label{sca_problem}
	\vspace{2.1pt}
	\begin{aligned}
		\hat{\bf{m}}^{\tau}=\min_{\bf{m}} &~~ \hat{f}_0^\tau(\bf{m}) \\
		\textup{s.t.}& ~~ \hat{f}_1^\tau(\bf{m}) \leq \bf{p}^{\textup{max}},\\
		&~ ~\sum_{n=1}^{N}m_n=1,\\
		& ~~0 \leq m_n \leq 1 , \forall n.
	\end{aligned}    
\end{equation}
If problem \eqref{sca_problem} turns out to be infeasible, the optimal solution $\hat{\bf{m}}^{\tau}$ is obtained by solving the following feasibility problem,
\begin{equation}\label{sca_problem_feasible}
	\begin{aligned}
		\hat{\bf{m}}^{\tau}=\min_{\bf{m},\boldsymbol{\mu}} &~~ \boldsymbol{\mu}\\
		\textup{s.t.}& ~~ \hat{f}_1^\tau(\bf{m}) \leq \bf{p}^{\textup{max}}+\boldsymbol{\mu},\\
		&~ ~\sum_{n=1}^{N}m_n=1,\\
		& ~~0 \leq m_n \leq 1 , \forall n.
	\end{aligned}    
\end{equation}
After solving for $\hat{\bf{m}}^{\tau}$, the model assignment policy is updated as
\begin{equation}\label{z_update_rule} 
	\begin{aligned}
		\bf{m}^{\tau+1} = (1 - \gamma^{\tau}) \bf{m}^{\tau} + \gamma^{\tau} \hat{\bf{m}}^{\tau},
	\end{aligned}    
\end{equation}
where $\gamma^{\tau} \in (0, 1)$ satisfies $\lim_{\tau \rightarrow\infty}\gamma^{\tau}=0$, $\sum_{\tau=0}^{\infty}\gamma^{\tau}=\infty$, and $\sum_{\tau=0}^{\infty}(\gamma^{\tau})^2<\infty${\color{black}\cite{liu2019stochastic}}.
{\color{black} To facilitate practical implementation and reproducibility, we provide explicit heuristics for choosing the hyperparameters $\rho^\tau $and $\gamma^\tau$. We suggest setting these hyperparameters as $\rho^\tau = \frac{1}{(\tau+1)^{\alpha}}$ and $\gamma^\tau = \frac{c}{\tau + c^{\prime}}$, where the parameter $\alpha$ typically ranges from 0.5 to 1 to satisfy the convergence conditions outlined in Lemma \ref{convergence_surrogate_function}. In practice, setting $\alpha \approx 0.8$, $c = 15$, and $c^{\prime} = 14$ has been found through simulations to achieve an effective balance between convergence speed and numerical stability.}

\begin{algorithm}[tbp]
	%	\vspace{5pt}
	%    \SetAlgoNoLine  %去掉之前的竖线
	\caption{Mixed-Timescale Model Assignment and Transceiver Optimization Algorithm} \label{alg_1}
	\mbox{\textbf{Initialize:} Model assignment policy $\bf{m}^{0}$, iteration index} $\tau=0$, and convergence tolerance $\epsilon$\;
	\textbf{\mbox{Step 1 (long-term model assignment optimization at the} beginning of inference task)} \\
	
	\Repeat{$\|  \bf{m}^{\tau} - \bf{m}^{\tau-1}\|  \leq \epsilon$}
	{
		
		\mbox{Obtain a channel sample $\bf{h}^{\tau}=\{\bf{h}_1^{\tau},\ldots,{\bf{h}_N^{\tau}}\}$ and cal-} \mbox{culate the short-term transceiver variables $\!\mathbf{a}^{*}(\bf{m}^\tau)$,} $\lbrace b_n^{*}(\bf{m}^\tau) \rbrace$ by solving the short-term problem $\ca{P}_s$\;
		
		\mbox{Update the surrogate functions $\hat{f}_i^{\tau}(\bf{m})$ according to \eqref{surrogate_obj_function}} and \eqref{surrogate_constraint_function}\;
		
		\uIf{problem \eqref{sca_problem} is feasible}{
			Solve problem \eqref{sca_problem} to obtain the optimal $\hat{\bf{m}}^{\tau}$
		}
		\Else{
			Solve problem \eqref{sca_problem_feasible} to obtain the optimal $\hat{\bf{m}}^{\tau}$
		}
		
		Update $\bf{m}^{\tau}$ according to \eqref{z_update_rule}\;
		\mbox{$\tau \leftarrow \tau + 1$\;}
	}
	
	\textbf{\mbox{Step 2 (short-term transceiver optimization at each all-} reduce step):} \\
	\mbox{Obtain the channel condition $\bf{h}$, and apply the short-term} \mbox{algorithm to solve the optimal transceiver variables with} the determined model assignment policy $\bf{m}$.
	
\end{algorithm}

\begin{figure}[!tb]
	\renewcommand\figurename{\small Fig.}
	\centering \vspace*{8pt} \setlength{\baselineskip}{10pt}
	\includegraphics[width = 0.5\textwidth,center,trim=5 83 60 40,clip]{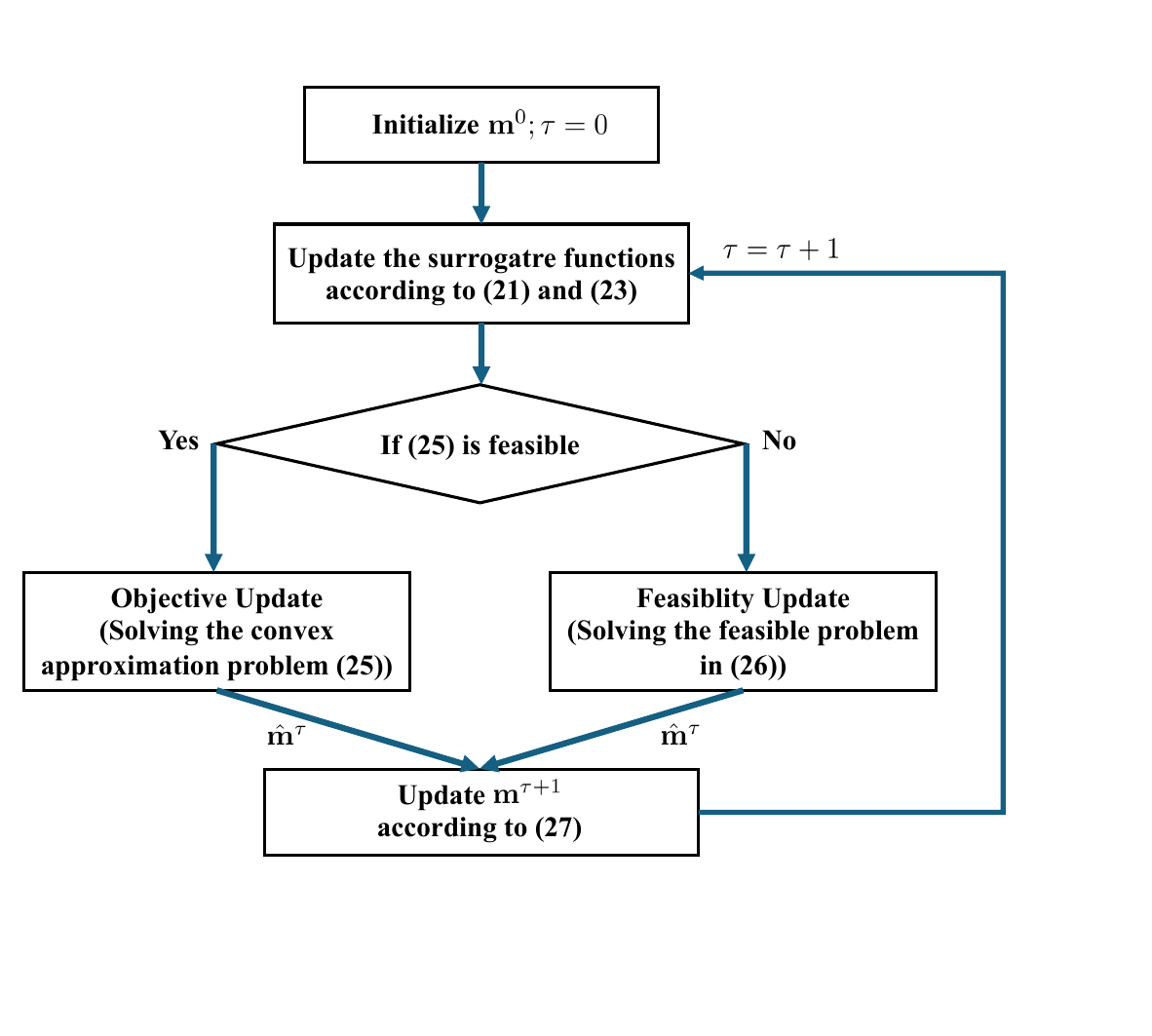}
	\caption{Block diagram of Algorithm 1}\label{fig_2}
\end{figure}

The above two steps iterate until convergence, i.e., $\|  \bf{m}^{\tau} - \bf{m}^{\tau-1}\|  \leq \epsilon$, where $\epsilon$ is the convergence tolerance.
The overall algorithm is outlined in Algorithm \ref{alg_1}, and the block diagram of the proposed algorithm is illustrated in Fig. \ref{fig_2}.

\begin{rem}
	{\color{black}The computational complexity of solving the short-term transceiver optimization problem $\ca{P}_s$ is at most $\ca{O}\left( N N_r^3 + N^2 N_r^2 + N^3\right) $, and is usually much lower in practice \cite[Theorem 3.12]{bomze2010interior}.
	Moreover, the most computation-expensive steps for the long-term optimization problem $\mathcal{P}_l$ are solving the constructed convex quadratic approximation problems \eqref{sca_problem} and \eqref{sca_problem_feasible}. Specifically, the computational complexity of solving problems \eqref{sca_problem} and \eqref{sca_problem_feasible} is at most in the order of $\ca{O}\left( 2N^4 + N^3\right) $. Then, the total computational complexity of Algorithm \ref{alg_1} is given by $\ca{O}\left( \tau^{\max} \left( N N_r^3 + N^2 N_r^2 + 2N^4+2N^3\right) \right) $, where $\tau^{\max}$ is the maximum iteration number of Algorithm \ref{alg_1}.}
%	We note that the optimization of model assignment policy only needs to be performed once for a given inference task, since devices should upload their corresponding model segments at the beginning of inference. Hence, Algorithm \ref{alg_1} has an affordable computational complexity.
	{\color{black} Both optimization and model assignment are performed only once at the beginning of the inference (or after substantial changes in channel or device conditions). Hence, while solving the optimization and loading model segments introduce a non-negligible one-time cost, the subsequent benefits from parallelized forward computation significantly outweigh this initial overhead, resulting in increased inference speed in practical settings.}
\end{rem}

{\color{black}
\begin{rem}
	In the proposed framework, the edge server, which possesses limited but non-negligible computational and memory resources, collects the channel state information and device capability information. Leveraging these data, the server solves the mixed-timescale optimization problem and assigns each participating edge device its respective portion of the model parameters.
\end{rem}}

\subsection{Convergence Analysis}

In this subsection, we analyze the asymptotic convergence performance of Algorithm \ref{alg_1} to a stationary point of the original problem $\ca{P}_1$.

We first show the convergence of the surrogate functions in the following lemma.

\begin{lem}\label{convergence_surrogate_function}
	Consider a sequence $\{\bf{m}^{\tau}\}_{\tau=0}^{\infty}$ converging to a limiting point $\bf{m}^{*}$, and define
	\begin{align}
		\vspace{2.1pt}
		\begin{split}\label{true_surrogate_func}
			\hat{f}_i(\bf{m}) = f_i(\bf{m}) + \nabla f_i(\bf{m}^*)^{\s{T}}\left( \bf{m} - \bf{m}^{*} \right)  + \eta_i \left\|  \bf{m} - \bf{m}^{*} \right\|^2,
		\end{split}
	\end{align}
	which satisfies $\hat{f}_i(\bf{m}^*) = f_i(\bf{m}^*) $ and $\nabla\hat{f}_i(\bf{m}^*) = \nabla f_i(\bf{m}^*) , \forall i\in \{0,1 \}$. Then, if the algorithm parameter $\rho$ satisfies $\sum_{\tau=0}^{\infty} \rho^{\tau}\tau^{-1/2}<\infty$, we have
	\begin{align}
		\vspace{2.1pt}
		\begin{split}\label{cvg_stochastic_surrogate_func}
			\lim_{\tau \to \infty}  \hat{f}_i^{\tau}(\bf{m}) = \hat{f}_i(\bf{m}),
		\end{split}
	\end{align}
	almost surely{\color{black}\cite{liu2019stochastic}}.
\end{lem}
\begin{proof}
	{The proof is presented in Appendix A.}
\end{proof}

To elaborate the convergence result, we need to introduce the Slater's condition for the converged surrogate function in the following.
\begin{defi}
	(Slater's Condition)
	Given the sequence $\{\bf{m}^{\tau}\}_{\tau=1}^{\infty}$ converging to a limiting point $\bf{m}^{*}$ and let $\hat{f}_1(\bf{m})$ be the converged surrogate function as defined in \eqref{true_surrogate_func}. The Slater's condition holds at $\bf{m}^{*}$ if there exists a constant $\bf{m} $ such that
%	\vspace{2.1pt}
	\begin{equation}
%		\vspace{2.1pt}
		\begin{aligned}
			&\hat{f}_1(\bf{m}) < \bf{p}^{\textup{max}},\\
			&\sum_{n=1}^{N}m_n=1,\\
			&0 \leq m_n \leq 1, \forall n.
		\end{aligned}    
	\end{equation}
\end{defi}
The Slater's condition is widely used in constrainted optimization algorithms (e.g., the majorization-minimization algorithm \cite{razaviyayn2014successive} and virtual queue-based online convex optimization algorithm \cite{zhang2023online}).
With Lemma \ref{convergence_surrogate_function} and the Slater's condition, we are ready to show the main convergence result of Algorithm \ref{alg_1} in the following theorem.
\begin{thm}\label{main_convergence}
	Let $\{\bf{m}^{\tau}\}_{\tau=1}^{\infty}$ denote the sequence of model assignment policies generated by Algorithm \ref{alg_1}.
	If the Slater's condition is satisfied at the limiting point $\bf{m}^{*}$ of the sequence $\{\bf{m}^{\tau}\}_{\tau=1}^{\infty}$, then $\bf{m}^{*}$ is a stationary point of problem $\ca{P}_1$ almost surely.
\end{thm}
\begin{proof}
	{The proof is presented in Appendix B.}
\end{proof}
In Section V-B, we further verify the convergence of Algorithm \ref{alg_1} through simulations.

\section{Extension to Multi-Antenna Devices}\label{sec_4}

In the previous sections, we analyzed the scenario of single-antenna edge devices to establish foundational insights for optimizing communication efficiency in distributed LLM inference.
In this section, we extend the proposed framework and algorithms to the multi-antenna setting.
By leveraging spatial multiplexing, the multi-antenna configuration further enhances communication efficiency and reduces inference latency, providing a more general and scalable solution.

\subsection{Problem Formulation}

Building upon the single-antenna setting, we now consider a more generalized scenario where edge devices in the distributed LLM inference system are equipped with multiple antennas.
Thus, the spatial diversity and spatial multiplexing are utilized to further improve communication efficiency.
Specifically, we consider the server and each edge device are equipped with $N_r$ and $N_t$ antennas, respectively.
Similar to the single-antenna case, all devices simultaneously upload their intermediate layer outputs through wireless multiple-access channels.
Let $\bf{z}_n=[z_{n,1}, \ldots, z_{n,L}]^{\s{T}}$ denote the per-round transmitted $L$ entries of device $n$'s intermediate output $\bf{Z}_n$.
Let $\mathbf{A} \in \mathbb{C}^{N_r \times L}$ and $\mathbf{B}_n \in \mathbb{C}^{N_t \times L}$ denote the aggregation beamforming matrix at the edge server and the data precoding matrix at device $n$, respectively.
Then, the received signal at the server after the AirComp can be derived as follows,
\begin{equation}\label{vector_z}
%	\vspace{2.1pt}
	\begin{aligned}
		&\hat{\bf{z}}  = \bf{A}^{\s{H}}\sum_{n=1}^{N}\mathbf{H}_n \mathbf{B}_n \bf{z}_n + \bf{A}^{\s{H}} \bf{n},
	\end{aligned}    
\end{equation}
where $\bf{H}_n\in \b{C}^{N_r\times N_t}$ denotes the uplink MIMO channel from device $n$ to the edge server.
{\color{black} In the multi-antenna setting, each device employs the precoding (beamforming) matrix to map its transmitted vector $\mathbf{z}_n$ onto multiple antennas for simultaneous transmission.}
The distortion of $\hat{\bf{z}}$ with respect to the desired target vector $\bf{z}  = \sum_{n=1}^{N} \bf{z}_n$ is measured by the MSE, defined as
%\vspace{2.1pt}
\begin{equation}\label{vector_mse}
%	\vspace{2.1pt}
	\begin{aligned}
		\textup{MSE}(\hat{\bf{z}},\bf{z}) = \mathbb{E}\left[(\hat{\bf{z}}-\bf{z})^{\s{H}}(\hat{\bf{z}}-\bf{z})\right].
	\end{aligned}    
\end{equation}
By substituting \eqref{vector_z} into \eqref{vector_mse}, the MSE can be explicitly represented as a function of transceiver beamforming matrices as follows,
\begin{equation}
	\begin{aligned}
		&\textup{MSE}(\mathbf{A},\left\lbrace \mathbf{B}_n  \right\rbrace ) \\
		&\!= \sum_{n=1}^{N} \textup{tr}\! \left( \! \left( \mathbf{A}^{\!\s{H}}  \mathbf{H}_n \mathbf{B}_n - \mathbf{I} \right) \! \left( \mathbf{A}^{\!\s{H}}  \mathbf{H}_n \mathbf{B}_n - \mathbf{I} \right)^{\s{H}} \right) \!+\! \sigma_z^2 \textup{tr}\left( \mathbf{A}^{\!\s{H}} \mathbf{A} \right).   \\
	\end{aligned}    
\end{equation}

To effectively utilize the heterogeneous computational capabilities of edge devices and mitigate communication distortions, we similarly investigate a joint model assignment and transceiver optimization problem.
Specifically, the joint optimization problem in the multi-antenna scenario can be formulated as follows,
\begin{equation}
%	\vspace{2.1pt}
	\begin{aligned}
		\ca{P}_2:~\min_{\bf{m}} &~~ \mathbb{E}_{\mathbf{H}}\left[ \min_{\mathbf{A},\left\lbrace \mathbf{B}_n  \right\rbrace} \textup{MSE}(\mathbf{A},\left\lbrace \mathbf{B}_n  \right\rbrace )    \right]  \\
		\textup{s.t.}& ~ ~e_n m_n s^{\textup{tot}} + \frac{L_0}{L} \textup{tr}\left( \mathbf{B}_n \mathbf{B}_n^{\s{H}} \right) \leq P_{n}^{\textup{max}}, \forall n,\\
		&~ ~\sum_{n=1}^{N}m_n=1,\\
		& ~~0 \leq m_n \leq 1 , \forall n,
	\end{aligned}    
\end{equation}
where the expectation $\mathbb{E}_{\mathbf{H}}\left[ \cdot \right] $ is taken over all random channel realizations $\mathbf{H} = \left\lbrace \bf{H}_n \right\rbrace_{n=1}^N $.

\subsection{Algorithm Development}

In this subsection, we extend Algorithm \ref{alg_1} to a more general case involving multi-antenna edge devices. Similarly, we first decompose problem $\ca{P}_2$ into a family of short-term subproblems and a long-term subproblem as follows.

\subsubsection{Short-term transceiver optimization for given model assignment policy $\bf{m}$ and channel condition $\bf{H}$}
%\vspace{2.1pt}
%\vspace{-1.5pt}
\begin{equation}
%	\vspace{2.1pt}
%	\vspace{-1.5pt}
	\begin{aligned}
		\ca{P}_{s}: \min_{\mathbf{A},\left\lbrace \mathbf{B}_n  \right\rbrace}  &~ \textup{MSE}(\mathbf{A},\left\lbrace \mathbf{B}_n  \right\rbrace )   \\
		\textup{s.t.}~~& ~ e_n m_n s^{\textup{tot}} + \frac{L_0}{L} \textup{tr}\left( \mathbf{B}_n \mathbf{B}_n^{\s{H}} \right) \leq P_{n}^{\textup{max}}, \forall n.
	\end{aligned}    
\end{equation}

\subsubsection{Long-term model assignment optimization based on the optimal solution $\mathbf{A}^{*}(\bf{m}),\left\lbrace \mathbf{B}_n^{*}(\bf{m})  \right\rbrace$ to problem $\ca{P}_{s}$}
%\vspace{2.1pt}
\begin{equation}
%	\vspace{2.1pt}
%\vspace{-1.5pt}
	\begin{aligned}
		\ca{P}_l:\min_{\bf{m}} &~~ \mathbb{E}_{\mathbf{H}}\left[ \textup{MSE}(\mathbf{A}^{*}(\bf{m}),\left\lbrace \mathbf{B}_n^{*}(\bf{m})  \right\rbrace )    \right]  \\
		\textup{s.t.}& ~~ e_n m_n s^{\textup{tot}} + \frac{L_0}{L} \textup{tr}\left( \mathbf{B}_n^{*}(\bf{m}) \mathbf{B}_n^{*}(\bf{m})^{\s{H}} \right) \leq P_{n}^{\textup{max}}, \forall n, \\
		&~ ~\sum_{n=1}^{N}m_n=1,\\
		& ~~0 \leq m_n \leq 1 , \forall n .
	\end{aligned}    
\end{equation}

To solve the short-term problem $\ca{P}_s$, we first simplify it by demonstrating that the zero-forcing (channel inversion) precoder is optimal conditioned on the aggregation beamformer.

\begin{lem}\label{lem_opt_precoder}
	For a given aggregation beamformer $\bf{A}$, the transmission MSE is minimized by using the zero-forcing precoders as follows,
%	\vspace{-3pt}
%\vspace{2.1pt}
	\begin{equation}\label{opt_precoder}
%		\vspace{2.1pt}
%		\vspace{-1pt}
		\begin{aligned}
			\mathbf{B}_n^* =\left(  \mathbf{A}^{\!\s{H}}  \mathbf{H}_n\right)^{\!\s{H}} \left(  \mathbf{A}^{\!\s{H}}  \mathbf{H}_n \mathbf{H}_n^{\s{H}} \mathbf{A} \right)^{\! -1}, \forall n.
		\end{aligned}    
	\end{equation}
\end{lem}
\begin{proof}
	{The proof of Lemma \ref{lem_opt_precoder} is similar to that of Lemma \ref{lem_opt_precoder_s} and thus omitted for brevity.}
\end{proof}

%\vspace{-1.5pt}
Let $\bf{G}$ represent the normalized aggregation beamformer that satisfies $\textup{tr}(\bf{G}\bf{G}^{\s{H}})=1$, and consequently $\bf{A}=\sqrt{\alpha} \bf{G}$ with $\alpha$ denoting the norm of $\bf{A}$.
By employing \eqref{opt_precoder}, the problem $\ca{P}_{s}$ can be reformulated as follows,
%\vspace{-1.5pt}
\begin{equation}\label{p_s_1}
%	\vspace{2.1pt}
%	\vspace{-1.5pt}
	\begin{aligned}
		\min_{\alpha, \bf{G}} &~~ \alpha\\
		\textup{s.t.} & ~~ e_n m_n s^{\textup{tot}} +  \frac{L_0}{\alpha L} \textup{tr}\left( \left( \bf{G}^{\s{H}} \bf{H}_n \bf{H}_n^{\s{H}} \bf{G} \right)^{-1}  \right) \leq P_{n}^{\textup{max}}, \forall n,\\
		& ~~\textup{tr}\left( \bf{G} \bf{G}^{\s{H}}  \right) = 1.
	\end{aligned}  
\end{equation}
The problem \eqref{p_s_1} remains challenging to be solved due to its non-convex constraints involving the term $\textup{tr}( ( \bf{G}^{\s{H}} \bf{H}_n \bf{H}_n^{\s{H}} \bf{G} )^{-1}  )$. To address this issue, we develop a tractable approximation of the problem by employing the following inequality,
%\vspace{-20pt}
%\vspace{0pt}
%\vspace{2.1pt}
\begin{equation}\label{sdr_ineq}
%	\vspace{2.1pt}
	\begin{aligned}
			 \textup{tr}\left( \left( \bf{G}^{\s{H}} \bf{H}_n \bf{H}_n^{\s{H}} \bf{G} \right)^{-1}  \right) \leq \frac{L}{\lambda_{\min}\left( \bf{H}_n^{\s{H}} \bf{G} \bf{G}^{\s{H}} \bf{H}_n \right)},
		\end{aligned}  
\end{equation}
where the equality holds when the channel is well-conditioned, i.e., the singular values of $\bf{H}_n$ are identical. By utilizing \eqref{sdr_ineq}, we reformulate an approximated version of problem \eqref{p_s_1} as follows,
%\vspace{-2.5pt}
%\vspace{2.1pt}
\begin{equation}\label{p_s_2}
%	\vspace{2.1pt}
%	\vspace{-2.5pt}
	\begin{aligned}
		\min_{\alpha, \bf{G}} &~~ \alpha\\
		\textup{s.t.} & ~~ \frac{L_0}{\alpha \lambda_{\min}\left( \bf{H}_n^{\s{H}} \bf{G} \bf{G}^{\s{H}} \bf{H}_n \right)} \leq P_{n}^{\textup{max}}-e_n m_n s^{\textup{tot}}, \forall n,\\
		& ~~\textup{tr}\left( \bf{G} \bf{G}^{\s{H}}  \right) = 1.
	\end{aligned}  
\end{equation}
Then, by introducing a new variable ${\hat{\bf G}}=\bf{G} \bf{G}^{\s{H}}$, an equivalent formulation of problem \eqref{p_s_2} is obtained as follows,
%\vspace{-2.5pt}
%\vspace{2.1pt}
\begin{equation}\label{p_s_3}
%	\vspace{2.1pt}
%	\vspace{-2.5pt}
	\begin{aligned}
		\min_{\alpha, \hat{\bf{G}}} &~~ \alpha\\
		\textup{s.t.} & ~~ \frac{L_0}{\alpha \lambda_{\min}\left( \bf{H}_n^{\s{H}} \hat{\bf{G}} \bf{H}_n \right)} \leq P_{n}^{\textup{max}}-e_n m_n s^{\textup{tot}}, \forall n,\\
		& ~~\textup{tr}(\hat{\bf{G}} ) = 1,~ \textup{rank}(\hat{\bf{G}})=L, ~ \hat{\bf{G}}\succeq 0.
	\end{aligned}  
\end{equation}
We observe that the only non-convex constraint in problem \eqref{p_s_3} is $\textup{rank}(\hat{\bf{G}})=L$. Therefore, we remove this constraint to obtain a relaxed version of problem \eqref{p_s_3} as follows,
%\vspace{-2pt}
\vspace{2.1pt}
\begin{equation}\label{p_s_4}
	\vspace{2.1pt}
%	\vspace{-1pt}
	\begin{aligned}
		\min_{\alpha, \hat{\bf{G}}} &~~ \alpha\\
		\textup{s.t.} & ~~ \frac{L_0}{\alpha \lambda_{\min}\left( \bf{H}_n^{\s{H}} \hat{\bf{G}} \bf{H}_n \right)} \leq P_{n}^{\textup{max}}-e_n m_n s^{\textup{tot}}, \forall n,\\
		& ~~\textup{tr}(\hat{\bf{G}} ) = 1, ~ \hat{\bf{G}}\succeq 0.
	\end{aligned}  
  \end{equation}
%\begin{lem}
%	Problem \eqref{p_s_4} is a convex problem.
%\end{lem}
%\vspace{-8pt}
%\begin{proof}
%	{The proof is omitted due to space limitation.}
%\end{proof}
%\vspace{-2pt}
The problem \eqref{p_s_4} can be proved to be a convex problem.
After solving problem \eqref{p_s_4} using a convex solver (e.g., the CVX toolbox in MATLAB \cite{grant2014cvx}) and obtaining the globally optimal solution $\hat{\bf{G}}^*$, we apply the Gaussian randomization algorithm \cite{luo2010semidefinite} to map the solution to a feasible, near-optimal solution for the original non-convex problem.

Next, we solve the long-term model assignment problem $\ca{P}_l$. The proposed stochastic SCA algorithm, initially introduced for the single-antenna case in Section \ref{sec_3}-B, can be directly extended to the multi-antenna scenario without requiring further modifications.
For clearer algorithmic description, we first reformulate the long-term problem $\ca{P}_l$ into an equivalent form as follows,
\begin{equation}
	\vspace{-3pt}
	\begin{aligned}		
		\min_{\bf{m}} &~~ f_0(\bf{m})=\mathbb{E}_{\mathbf{H}}\left[ \textup{MSE}(\mathbf{A}^{*}(\bf{m}),\left\lbrace \mathbf{B}_n^{*}(\bf{m})  \right\rbrace )    \right]  \\
		%		\vspace{-10pt}
		\textup{s.t.}& ~~ f_1(\bf{m})= s^{\textup{tot}} \textup{diag}(\bf{e} \bf{m}^{\s{T}})  + \frac{L_0}{L} \bf{e}_c\left(\bf{m} \right) \leq \bf{p}^{\textup{max}},\\
		&~~ \bf{m}^{\textsf{T}}\boldsymbol{1}=1, \bf{m}\geq 0,
	\end{aligned}    
\end{equation}
where 
\vspace{-2pt}
$$\bf{e}_c\!\left(\bf{m} \right) \!=\! [\textup{tr}\!\left( \mathbf{B}_1^{*}(\bf{m}) (\mathbf{B}_1^{*}(\bf{m}))^{\s{H}} \right)\!, \!\ldots\!, \textup{tr}\!\left( \mathbf{B}_N^{*}(\bf{m}) (\mathbf{B}_N^{*}(\bf{m}))^{\s{H}} \right)]^{\s{T}}\!,$$
$\bf{p}^{\textup{max}}=[{P}_1^{\textup{max}}, \ldots, {P}_N^{\textup{max}}]^{\s{T}}$, and $\bf{e} = [e_1,\ldots,e_N]^{\s{T}}$.
The main structure of the proposed stochastic SCA algorithm remains intact, and it iteratively performs the following two steps: First, quadratic surrogate functions $\hat{f}_0(\bf{m})$, $\hat{f}_1(\bf{m})$ are constructed to approximate the non-convex components of the original objective and constraint functions $f_0(\bf{m})$, $f_1(\bf{m})$, respectively.
Then, the resulting convex quadratic approximation problem is solved, and the long-term model assignment policy is updated based on the solution.
Here, we omit the details of these two steps for brevity. In Section \ref{sec_5}-B, we also demonstrate the convergence of the proposed algorithm for the multi-antenna scenario through simulations.

\section{Simulation Results}\label{sec_5}

%In this section, we evaluate the performance of the proposed air all-reduce algorithm by conducting simulations on real-world datasets using the state-of-the-art LLMs.

%\vspace{-20pt}
%\vspace{-2pt}
\subsection{Simulation Setups}

\subsubsection{LLM Inference Model Setting}
{\color{black}All simulations are performed on a desktop server equipped with Nvidia GeForce RTX 4070Ti GPU and Intel Core i9 CPU, using PyTorch 2.0 with CUDA 11.7. We set up $N$ virtual machines (VMs) with each VM simulating a distinct edge device.
Each VM is allocated 4 CPU cores, 16 GB RAM, and 128 GB storage space, ensuring efficient utilization of computational resources and optimized parallel processing.}
For evaluation, we utilize the LLaMA2 \cite{touvron2023llama} and LLaMA3 \cite{dubey2024llama} models due to their state-of-the-art performance among open-source models. Additionally, we employ the WikiText-2 dataset \cite{merity2017regularizing}, which is widely used in the field of LLM inference for benchmarking and evaluation purposes.
{\color{black}We have released our implementation on GitHub: \url{https://github.com/zklasd24/distributed_llama_AirComp}, which builds upon the open-source project \textit{Distributed Llama}~\cite{dllama}.}

The primary performance metric for inference accuracy is perplexity \cite{alon2023detecting}, which is a widely recognized measure of a LLM's capability to predict the next word in a sequence. It is defined mathematically as follows,
\begin{equation} 
	\vspace{2.1pt}
	\begin{aligned} 
		&\mathrm{Perplexity} =\exp \!\left(\!-\frac{1}{L_{\textup{txt}}} \sum_{k=1}^{L_{\textup{txt}}} \log \b{P}{}\left(w_k \!\mid\! w_1, \ldots, w_{k-1}\right)\right)\!,
		\end{aligned}
\end{equation}
where $\b{P}{}\left(w_k \mid w_1, \ldots, w_{k-1}\right)$ denotes the model's predicted probability for the next word $w_k$, and $L_{\textup{txt}}$ is the text length.
Lower perplexity values indicate better inference performance, reflecting the model’s accuracy in generating subsequent tokens.

\subsubsection{Communication Model Setting}

The number of antennas at the edge server is $N_r=16$, and each edge device has single antenna or $N_t=4$ antennas for different cases. The bandwidth between the edge server and edge devices is $B = 10$ MHz.
The uplink channels are assumed to be independent and identically distributed (i.i.d.) Rician fading \cite{tse2005fundamentals}, modeled as i.i.d. complex Gaussian random variables with non-zero mean $\mu = 1$ and variance $\sigma^2 = 1$. Moreover, the maximum power budget is set as $P_{n}^{\textup{max}}= 10$ and the noise variance at the edge server is assumed to be 1.

\subsection{Algorithm Convergence}

\begin{figure}[!tb]
	\renewcommand\figurename{\small Fig.}
	\hspace{5pt} \setlength{\baselineskip}{10pt}
	\includegraphics[width = 0.53\textwidth,center,trim=0 0 0 20,clip]{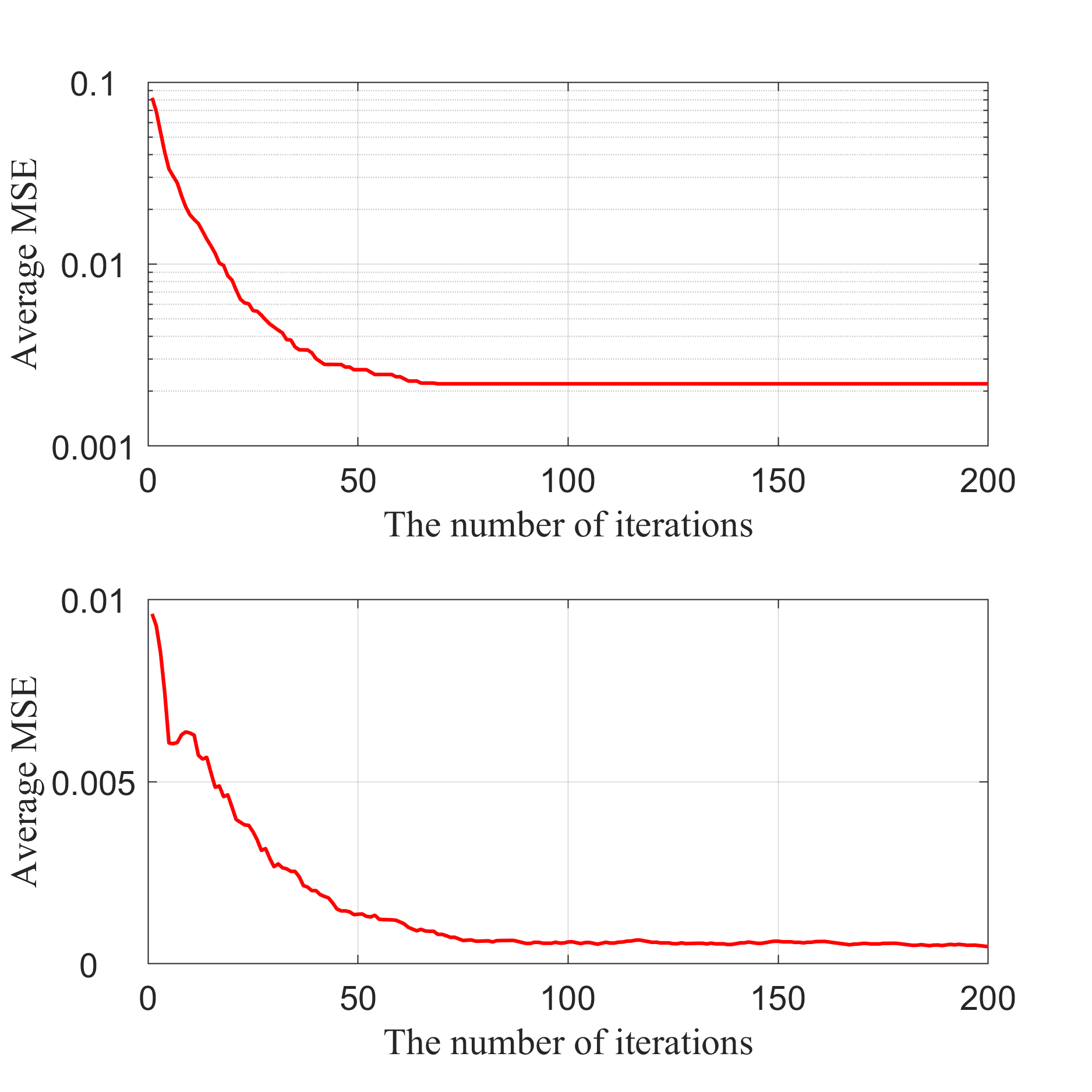}
	%	\vspace{-15pt}
	\caption{\color{black}Convergence of Algorithm 1 for the scenarios of single-antenna devices (Top) and multi-antenna devices (Bottom).}\label{fig_3}
\end{figure}

In this subsection, we analyze the convergence behavior of the propose d algorithm for both single-antenna and multi-antenna scenarios.
The parameter parameters are set as $\epsilon=0.001$, $\rho^{\tau}=[1/((\tau+1)^{4/5})]$, and $\gamma^\tau=15/(14+\tau)$.
As illustrated in Fig. \ref{fig_3}, the proposed algorithm demonstrates rapid convergence, reaching a stationary point within approximately 100 iterations.
The swift convergence speed ensures that the distributed LLM inference system can quickly adapt to varying network conditions, enabling real-time inference especially in latency-sensitive applications.
Moreover, the consistent performance across both single-antenna and multi-antenna settings suggests the robustness of the proposed algorithm to various network scenarios.

\begin{figure*}[htbp]
	%		\vspace{1pt}
	\centering
	% 第一个子图
	\begin{subfigure}{0.328\textwidth}
		%		\centering
		\hspace{-4pt}
		\includegraphics[width=1.05\linewidth]{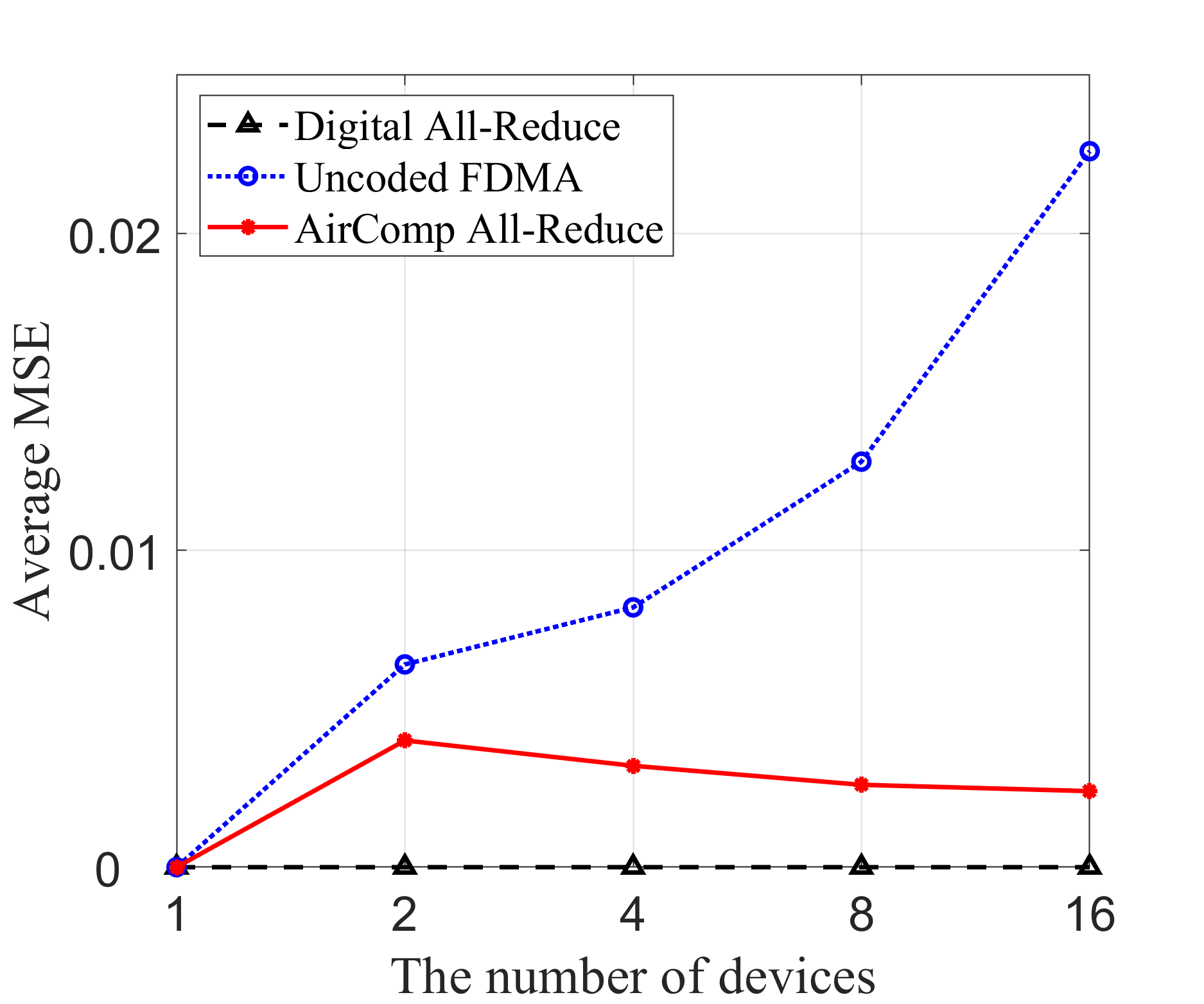}
		\caption{}
		\label{fig:subfig1}
	\end{subfigure}
	% 第二个子图
	\begin{subfigure}{0.328\textwidth}
		%		\centering
		\hspace{-4pt}
		\includegraphics[width=1.05\linewidth]{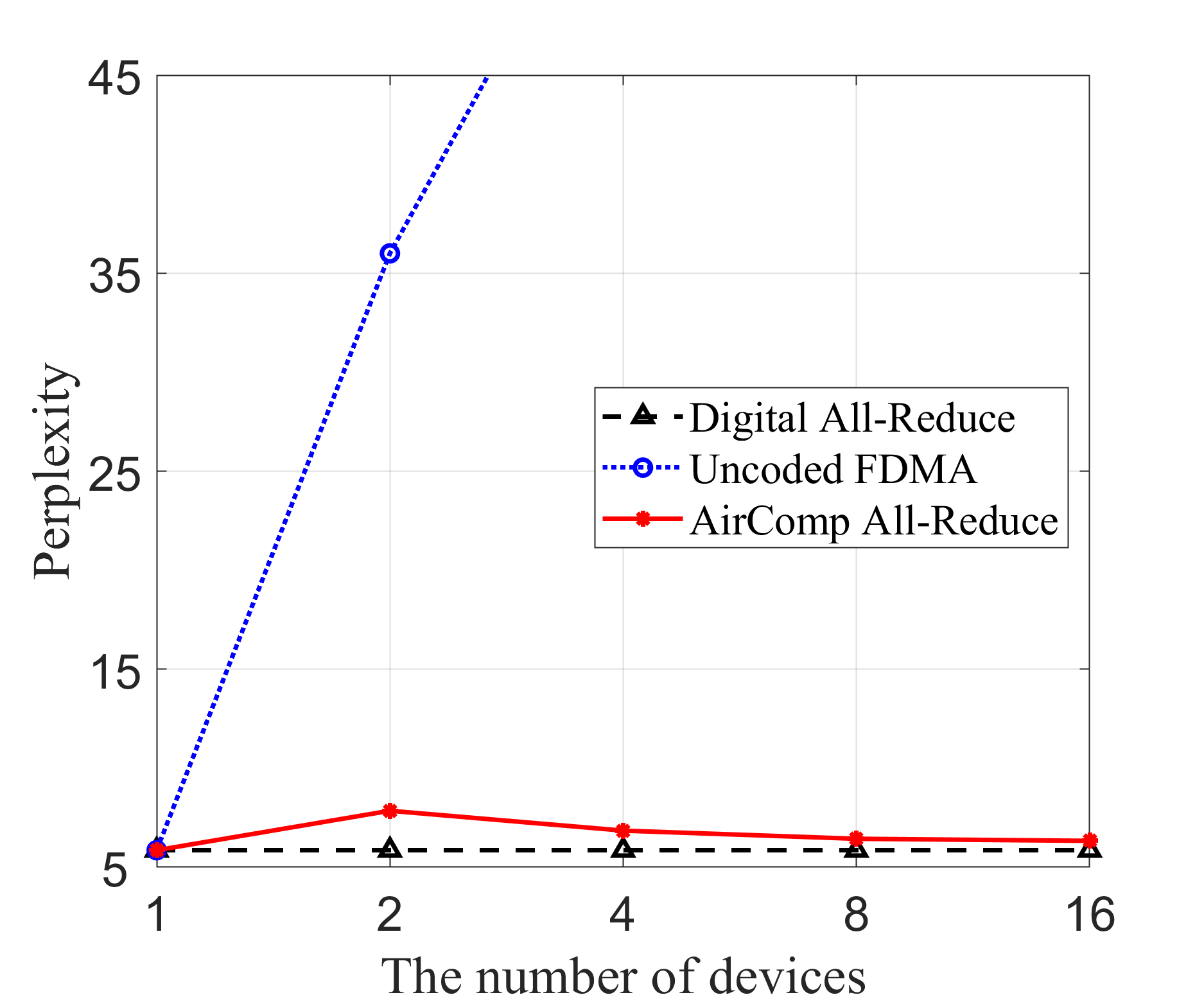}
		\caption{}
		\label{fig:subfig2}
	\end{subfigure}
	% 第三个子图
	\begin{subfigure}{0.328\textwidth}
		%		\centering
		\hspace{-4pt}
		\includegraphics[width=1.05\linewidth]{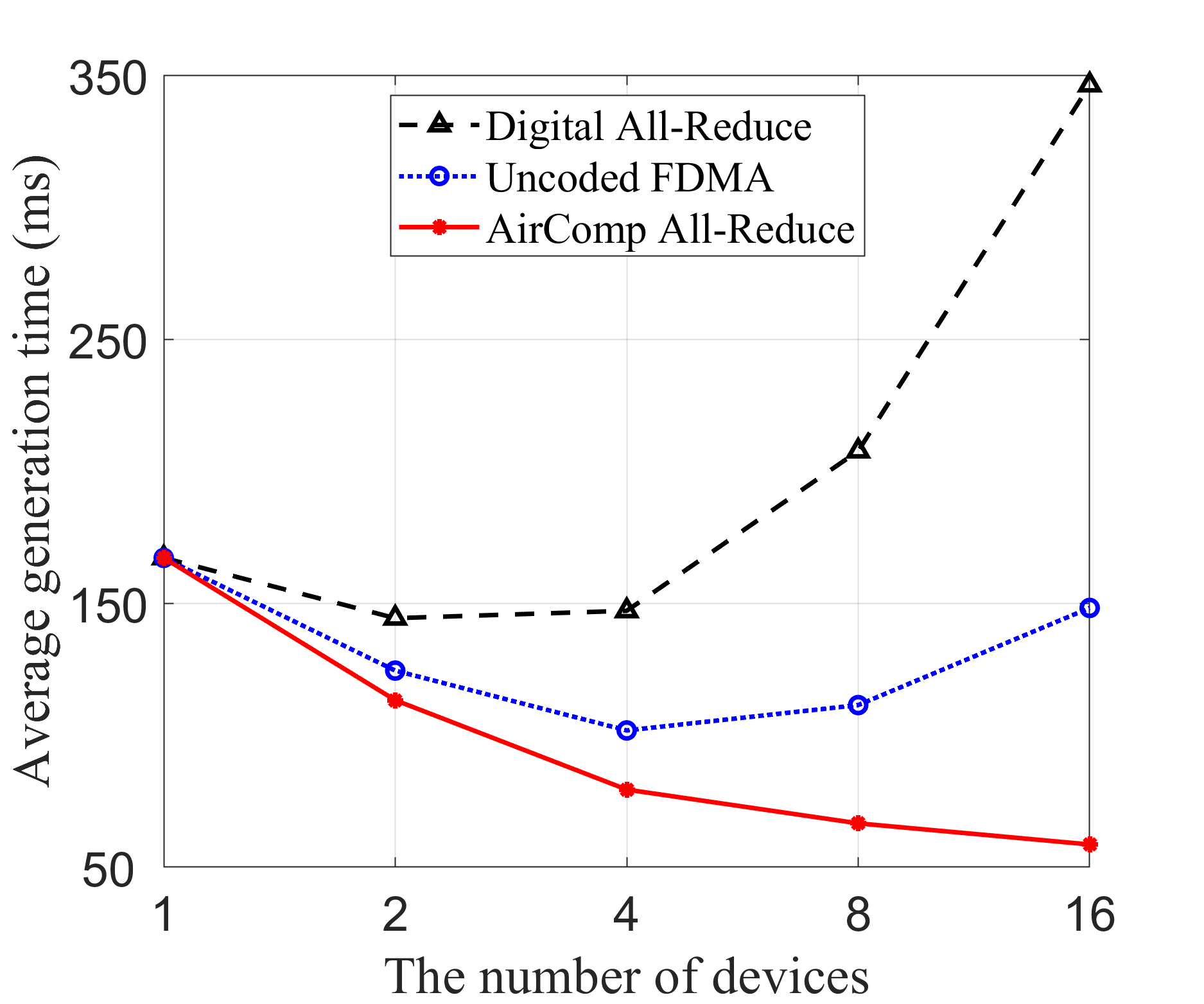}
		\caption{}
		\label{fig:subfig3}
	\end{subfigure}
	%	\vspace{-14pt}
	\caption{The average MSE (a), perplexity (b), and average generation time (c) versus the number of edge devices for the scenario of single-antenna d evices.}
	\label{fig:three_subfigs}
\end{figure*}

\subsection{Performance Evaluation}

In this subsection, we compare the performance of the proposed AirComp all-reduce approach with the following two benchmark schemes.

\begin{itemize}
	\item \textbf{Digital All-Reduce}: All devices upload intermediate layer outputs using a traditional broadband digital multiple-access scheme, with each transmitted symbol quantized to $Q=8$ bits. To prevent multi-user interference, orthogonal frequency division multiple-access (OFDMA) is employed, assigning each sub-channel to one device \cite{goldsmith2005wireless}.
	\item \textbf{Uncoded FDMA}: This scheme similarly employs the OFDMA technique, with each device occupying a dedicated sub-channel to upload intermediate layer outputs in an uncoded analog manner.
	
\end{itemize}

\setlength{\tabcolsep}{17.71pt}
\renewcommand{\arraystretch}{1.6} % 1.5倍默认行高
\begin{table}[t]
	\vspace{15pt}
	\centering
	\normalsize
	\begin{tabular}{cc}
		\toprule
		\multirow{1}{*}{\textbf{Transmission Scheme}}  &\multirow{1}{*}{\textbf{Transmission Time}}\\
		\midrule
		Digital All-Reduce & {\Large $\frac{N L_0Q}{B\log_2\left( 1 + \textup{SNR}_{\textup{rx}} N \right) }$}  \\
		\midrule
		Uncoded FDMA & {\Large $\frac{N L_0}{B}$}  \\
		\midrule
		AirComp All-Reduce & {\Large $\frac{L_0}{B}$} \\
		\bottomrule
		%		\vspace{0.1pt}
	\end{tabular}
	%	\vspace{-4pt}
	%	\vspace{-3pt}
	\caption{Transmission time for different transmission schemes.}\label{table_1}
	\vspace{-6pt}
\end{table}

\begin{figure*}[htbp]
	%	\vspace{-3pt}
	\centering
	% 第一个子图
	\begin{subfigure}{0.328\textwidth}
		%		\centering
		\hspace{-4pt}
		\includegraphics[width=1.05\linewidth]{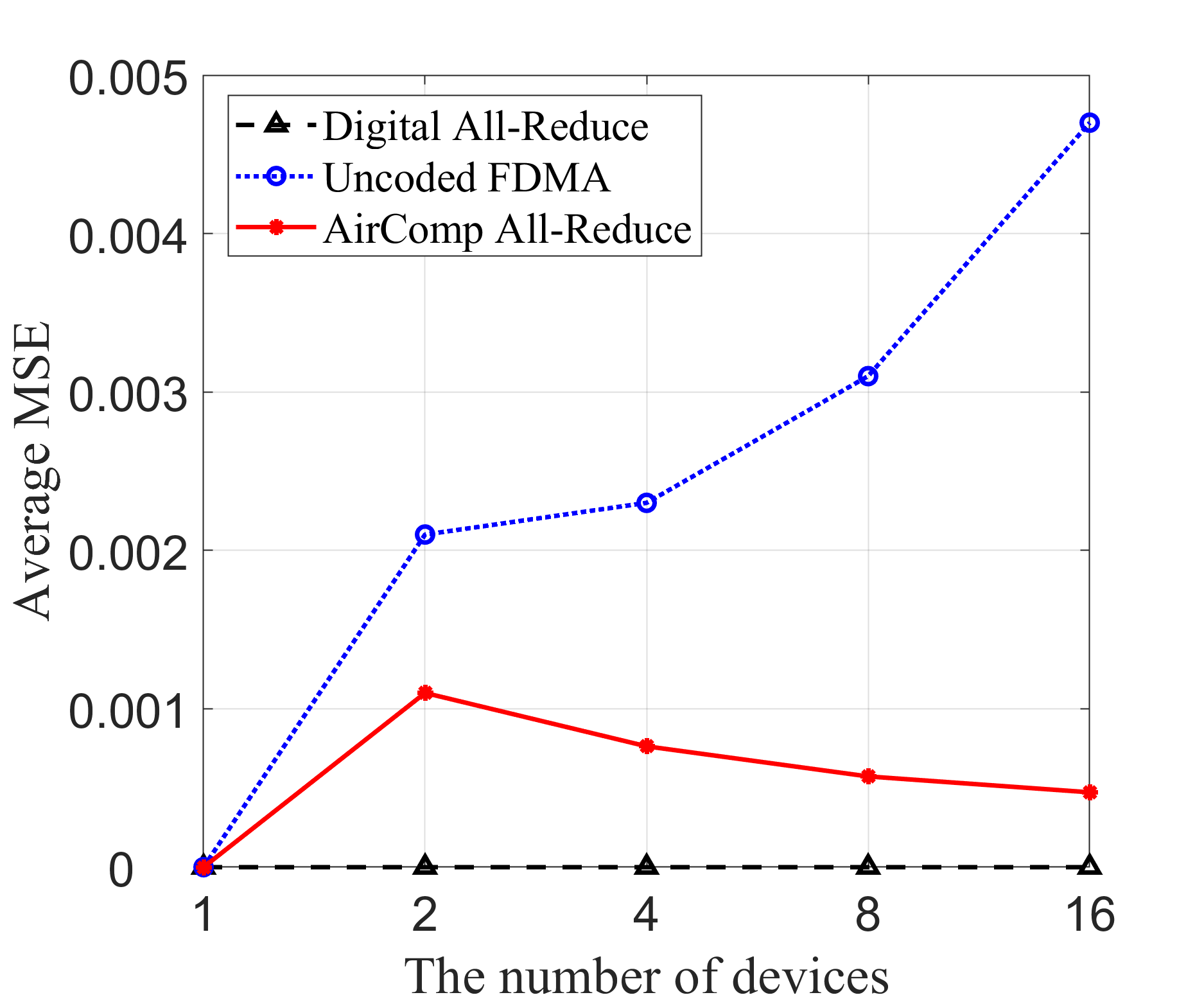}
		\caption{}
		\label{fig:subfig1_mimo}
	\end{subfigure}
	% 第二个子图
	\begin{subfigure}{0.328\textwidth}
		%		\centering
		\hspace{-4pt}
		\includegraphics[width=1.05\linewidth]{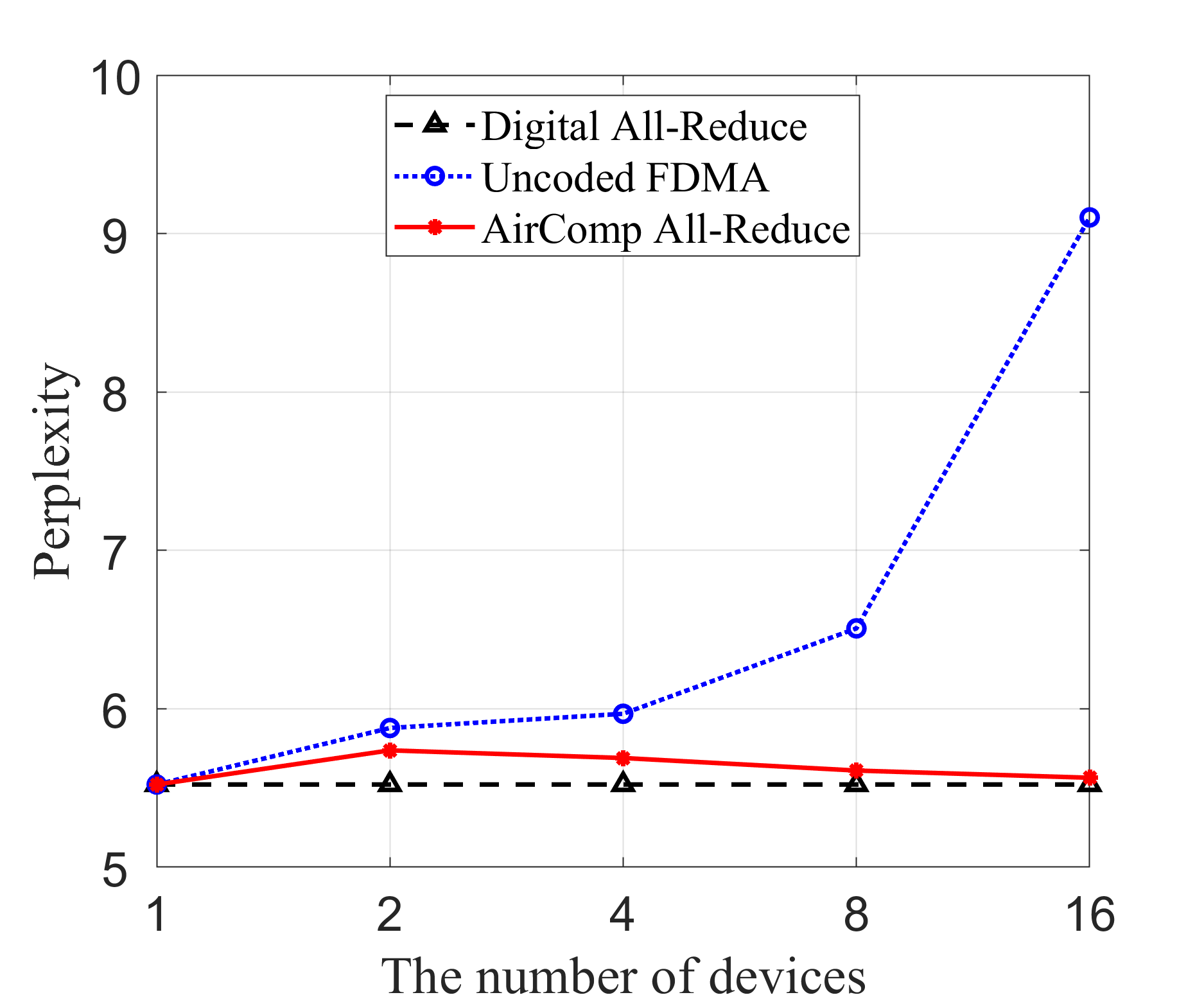}
		\caption{}
		\label{fig:subfig2_mimo}
	\end{subfigure}
	% 第三个子图
	\begin{subfigure}{0.328\textwidth}
		%		\centering
		\hspace{-4pt}
		\includegraphics[width=1.05\linewidth]{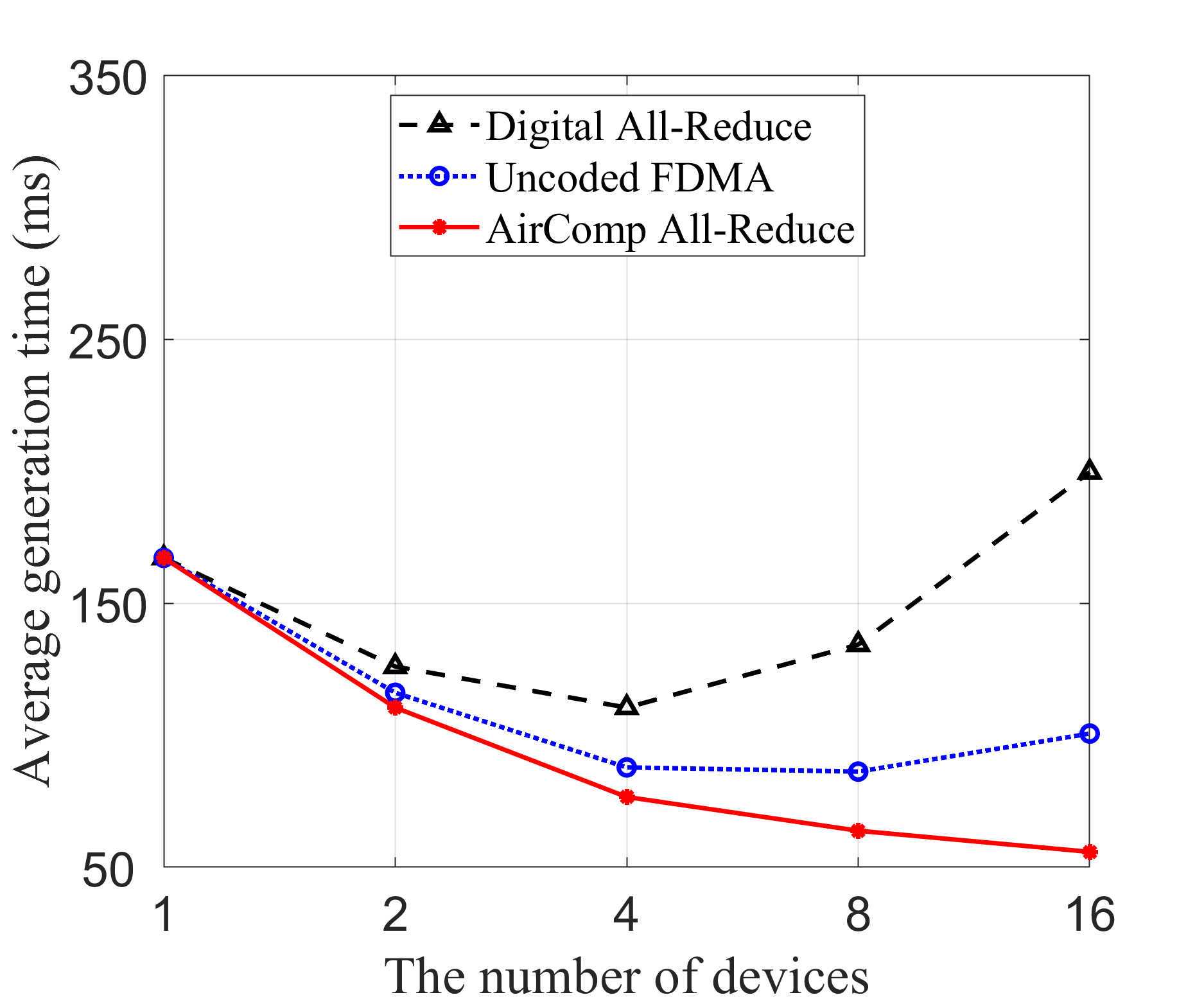}
		\caption{}
		\label{fig:subfig3_mimo}
	\end{subfigure}
		\vspace{-14pt}
	\caption{\color{black}The average MSE (a), perplexity (b), and average generation time (c) versus the number of edge devices for the scenario of multi-antenna devices.}
	\label{fig:three_subfigs_mimo}
	
	\vspace{4pt}
\end{figure*}

In Fig. \ref{fig:three_subfigs}, we compare the inference performance of different transmission schemes using the LLaMA3 model with 8 billion parameters, across three key performance metrics: transmission MSE, perplexity, and average generation time. 
In Fig. \ref{fig:three_subfigs}(a), the proposed AirComp all-reduce approach consistently achieves low MSE across all device counts, significantly outperforming the uncoded FDMA scheme, which exhibits a near-linear increase in MSE as the number of devices grows. The digital all-reduce method achieves near-zero MSE across all configurations. However, it has significantly higher communication latency.
In Fig. \ref{fig:three_subfigs}(b), perplexity follows the same trend as the transmission MSE. The AirComp all-reduce method maintains stable, low perplexity across all device configurations, while the perplexity of uncoded FDMA rises sharply with more devices. Digital all-reduce performs similarly to AirComp all-reduce, maintaining low perplexity.
Turning to the average generation time in Fig. \ref{fig:three_subfigs}(c), we observe a notable distinction among the three methods. Here, the total inference time is defined as the sum of local computation time and the time taken to transmit the local outputs.
The local computation time is obtained through experimental measurements, while the communication time is estimated based on different transmission methods as outlined in Table \ref{table_1}, where $\textup{SNR}_{\textup{rx}}$ denotes the average receive signal-to-noise ratio (SNR).
We observe that AirComp all-reduce consistently demonstrates the lowest latency, particularly as the number of edge devices grows.
The digital all-reduce scheme shows a significant increase in generation time with more devices due to increased communication overhead, while the uncoded FDMA method provides moderate improvements but still lags behind AirComp all-reduce.
{\color{black}The proposed AirComp all-reduce approach exhibits superior scalability compared to traditional communication strategies. Specifically, by exploiting analog signal superposition inherent in wireless channels, AirComp enables simultaneous aggregation of signals from multiple devices within a single communication slot. Consequently, unlike traditional communication schemes, whose overhead increases linearly with the number of participating devices, the AirComp all-reduce approach maintains low communication latency even as device count grows.}

Fig. \ref{fig:three_subfigs_mimo} expands on the simulation results by evaluating the performance of the proposed method in a more general setting of multi-antenna devices. In this scenario, the digital all-reduce scheme maintains the lowest MSE and perplexity. However, its average generation time grows considerably with an increasing number of devices, indicating scalability limitations in practice.
The proposed AirComp all-reduce scheme, while exhibiting a slight increase in MSE compared to digital all-reduce, remains competitive in terms of perplexity and demonstrates the shortest generation time across all configurations. This makes it an promising choice for applications where low latency is critical, and slight trade-offs in accuracy are acceptable.
On the other hand, the uncoded FDMA scheme's performance degrades significantly with more devices, reflected by steep increases in both MSE and perplexity.

\renewcommand{\arraystretch}{1.8} 
\setlength{\tabcolsep}{3pt}
\begin{table*}[t]
	\vspace{2pt}
	\centering
	\small
	\begin{tabular}{lcccccccccccccccc}
		\toprule
		&  \multicolumn{16}{c}{\textbf{{\normalsize Average generation time per token (ms)}}} \\
		\hhline{|~|-|-|-|-|-|-|-|-|-|-|-|-|-|-|-|-|}
		\multirow{1}{*}{\textbf{{\normalsize Model}}}  & \multicolumn{4}{c}{\textbf{{\normalsize LLaMA2-7B}}} & \multicolumn{4}{c}{\textbf{{\normalsize LLaMA2-13B}}} & \multicolumn{4}{c}{\textbf{{\normalsize LLaMA2-70B}}} & \multicolumn{4}{c}{\textbf{{\normalsize LLaMA3-70B}}}\\
		\multirow{1}{*}{\textbf{{\normalsize Device Number}}} &  \textbf{1} & \textbf{2} & \textbf{4} & \textbf{8} &  \textbf{1} & \textbf{2} & \textbf{4} & \textbf{8}& \textbf{1} & \textbf{2} & \textbf{4} & \textbf{8} &  \textbf{1} & \textbf{2} & \textbf{4} & \textbf{8} \\
		\midrule
		{\normalsize Digital All-Reduce} & 114.2 & 85.2 & 79.5 & 108.3 & 217.3 & 174.0 & 176.6  &261.4 &\hspace{4.2pt}$\textup{N/A}^{*}$&807.3&729.7&981.6&N/A&893.2&783.8&1033.6\\
		{\normalsize AirComp All-Reduce} & 114.2 & 69.7 & 45.7 & {\textbf{37.8}} & 217.3 & 128.5 & 81.3 & {\textbf{66.4}} & N/A & 660.9 &423.0& {\textbf{354.2}}&N/A&746.8&477.1& {\textbf{406.0}}\\
		\bottomrule
		\multicolumn{5}{c}{{\normalsize *: Not available due to insufficient memory.}}&
		%		\vspace{0.1pt}
	\end{tabular}
	%	\vspace{-4pt}
	%	\vspace{-3pt}
	\caption{Average generation time for different models across varying device numbers, with the shortest average generation time for each model being highlighted in bold.}\label{table_2}
	%	\vspace{-2pt}
\end{table*}

To further validate the effectiveness of the proposed algorithm, we conduct additional experiments using larger models, including LLaMA2 with 7, 13, and 70 billion parameters, and LLaMA3 with 70 billion parameters. In Table \ref{table_2}, it is observed that AirComp all-reduce method consistently demonstrates superior performance in terms of reduced generation time, particularly as the number of devices increases.
Across various device and model configurations, AirComp all-reduce achieves up to 4x faster generation speed, demonstrating its significant advantages for distributed LLM inference, especially with large-scale models.

Overall, the AirComp all-reduce approach emerges as a balanced and scalable solution, effectively managing the trade-offs between latency, accuracy, and scalability in both single-antenna and multi-antenna environments. This highlights its potential for deployment in practical, large-scale wireless scenarios.

{\color{black}
\subsection{Comparison with Centralized Inference Approach}
In this subsection, we compare the proposed AirComp-based distributed inference framework with the traditional centralized inference approach. Table \ref{table_cen_decen} compares the per-token generation latency for centralized versus distributed LLM inference across different large models.
As shown in the table, although the centralized inference does not incur a communication overhead, it suffers from significantly higher per-token computation time. In contrast, the proposed distributed inference approach partitions the model across multiple devices, substantially reducing each device’s computational load. Despite introducing modest communication overhead, the proposed distributed scheme achieves significantly lower total inference latency per token.
Hence, for large-scale LLMs with billions of parameters, distributing both the model storage and compute cost across multiple devices proves far more feasible and efficient than hosting the entire model on a single node.
Moreover, both per-token local computation time and communication overhead increase substantially as the number of transformer layers grows. However, it is noteworthy that the distributed inference approach consistently maintains a significant latency advantage over centralized inference across all models with different number of layers.}

\renewcommand{\arraystretch}{1.2} 
\setlength{\tabcolsep}{4pt}
\begin{table*}[t]
	\vspace{2pt}
	\centering
	\small
	\color{black}
	\begin{tabular}{|c|c|c|cc|c|}
		\toprule
		\multirow{1}{*}{\textbf{{  Model}}}  &  \multicolumn{1}{c|}{\textbf{  \makecell[c]{Number of  \\ Transformer\\ Layers }}} & \multicolumn{1}{c|}{\textbf{{  Method}}} & \multicolumn{1}{c}{\textbf{{  \makecell[c]{Per-Token Local \\ $~$Computation Time (ms)$~$}}}} & \multicolumn{1}{c}{\textbf{{  \makecell[c]{$~$Per-Token \\ $~$Communication $~$\\ Time (ms)}}}} & \multicolumn{1}{|c|}{\textbf{{  \makecell[c]{Per-Token Total \\ Generation Time (ms)}$~$}}}\\
		\midrule
		\multirow{2}{*}{{$~$   LLaMA2-7B}$~$} &  \multirow{2}{*}{{$~$   32}$~$} & Centralized & 114.2 & 0 & 114.2   \\
		 && $\quad$ Distributed $\quad$& 26.0 & 11.8 & 37.8 \\
		 \midrule
		 \multirow{2}{*}{{$~$   LLaMA2-13B}$~$} &\multirow{2}{*}{{$~$   40}$~$} &  Centralized & 217.3 & 0 & 217.3   \\
		 && $\quad$ Distributed $\quad$& 38.5 & 27.9 & 66.4 \\
		 \midrule
		 \multirow{2}{*}{{$~$   LLaMA2-70B}$~$} &\multirow{2}{*}{{$~$   80}$~$} &  Centralized & 1152.6 & 0 & 1152.6   \\
		 && $\quad$ Distributed $\quad$& 264.6 & 89.6 & 354.2 \\
%		 \midrule
%		 \multirow{2}{*}{{$~$ \normalsize LLaMA3-70B}$~$} &  Centralized & 1254.4 & 0 & 1254.4   \\
%		 & $\quad$ Distributed $\quad$& 316.3 & 89.7 & 406.0 \\
		\bottomrule
		%		\vspace{0.1pt}
	\end{tabular}
	%	\vspace{-4pt}
	%	\vspace{-3pt}
	\caption{\color{black}Comparison of Centralized v.s.\ Decentralized Inference across Different Models: Per-Token Computation, Communication, and Total Latency.}\label{table_cen_decen}
	%	\vspace{-2pt}
\end{table*}

%\vspace{-1pt}
\section{Conclusion}\label{sec_6}
%\vspace{-1.5pt}

In this paper, we proposed a novel distributed on-device LLM inference framework employing tensor parallelism.
To mitigate the communication overhead from frequent all-reduce steps in tensor parallelism, we proposed a communication-effcient AirComp all-reduce approach.
Moreover, to minimize the average transmission MSE, we formulated a joint model assignment and transceiver design problem, which can be derived as a mixed-timescale stochastic non-convex optimization.
We further developed an efficient two-stage algorithm that decomposed the original problem in short-term transceiver optimization and long-term model assignment optimization problems, which were solved by leveraging the SDR and stochastic SCA, respectively. 
We proved that the proposed algorithm can converge almost surely to a stationary point of the original problem.
Simulation results demonstrated that the proposed approach significantly reduced inference latency while improving inference accuracy, making distributed on-device LLM inference feasible for resource-constrained edge devices.

There are several promising directions for further advancing distributed on-device LLM inference systems.
{\color{black}One important research direction is experimentally validating the proposed AirComp-based distributed inference framework using real-world wireless hardware setups, further assessing practical performance and robustness. In addition, exploring cluster-based hierarchical AirComp designs and distributed transceiver optimization methods can effectively address potential scalability bottlenecks arising from synchronization overhead, channel estimation complexity, and computational demands in large-scale device networks.}

%\vspace{10pt}
\section*{Appendix}
\subsection{Proof of Lemma \ref{convergence_surrogate_function}}\label{app_a}

According to the assumption that channel statistics remain constant throughout the inference process, we have that the sample-wise approximation of the average MSE, $\bar{f}_0(\bf{m}^{\tau}) $, satisfying
\vspace{2.1pt}
\begin{equation}\label{proof_1_1}
	%	\vspace{-1.5pt}
	\begin{aligned}
		&\bar{f}_0(\bf{m}^{\tau})   \xrightarrow{\textup{~a.s.~}}   f_0(\bf{m}^{\tau}),
	\end{aligned}  
\end{equation}
\begin{equation}\label{proof_1_2}
	%	\vspace{-1.5pt}
	\begin{aligned}
		\mathbb{E}[\|  \bar{f}_0(&\bf{m}^{\tau})  - f_0(\bf{m}^{\tau}) \|] = \ca{O}\left( \frac{1}{\sqrt{\tau}} \right), 
	\end{aligned}  
\end{equation}
which follow from the law of large numbers and the central limit theorem, respectively. Then, combining \eqref{proof_1_1} and \eqref{proof_1_2} into \eqref{surrogate_obj_function}, we have
\vspace{2.1pt}
\begin{equation}\label{proof_1_3}
	\vspace{2.1pt}
	%	\vspace{-1.5pt}
	\begin{aligned}
		\lim_{\tau\to \infty}|  \hat{f}_i^{\tau}(&\bf{m}^{\tau})  - f_i(\bf{m}^{\tau}) | = 0, 
	\end{aligned}  
\end{equation}
for $i=0,1$.
Equation \eqref{proof_1_3} indicates the convergence of $\hat{f}_i^{\tau}$, and we then need to prove the convergence of $\nabla\hat{f}_i^{\tau}$ as follows,
\begin{equation}\label{proof_1_4}
	%	\vspace{-1.5pt}
	\begin{aligned}
		\lim_{\tau\to \infty}|  \nabla\hat{f}_i^{\tau}(&\bf{m}^{\tau})  - \nabla f_i(\bf{m}^{\tau}) | = 0. 
	\end{aligned}  
\end{equation}
It is easy to verify that the MSE funtion $\bar{f}_0$ and its derivation $\nabla\bar{f}_0$ are Lipschitz continuous, according to the fact that the channel sample is always bounded in practice.
Then, we can obtain that
	\vspace{2.1pt}
\begin{equation}\label{proof_1_5}
	\vspace{2.1pt}
	\begin{aligned}
		&\| \mathbb{E}[\bf{u}_0^{\tau}] - \nabla  f_0(\bf{m}^{\tau})  \|   \\
		\leq& \mathbb{E}[\|  \nabla\bar{f}_0(\bf{m}^{\tau}) -  \nabla  f_0(\bf{m}^{\tau})  \|]\\
		\aleq  & \ca{O}\left( \frac{1}{\sqrt{\tau}} \right),
	\end{aligned}  
\end{equation}
where (a) holds since $\nabla\bar{f}_0$ is Lipschitz continuous.
From $\sum_{\tau=0}^{\infty} \rho^{\tau}\tau^{-1/2}<\infty$, we can obtain that
\begin{equation}\label{proof_1_6}
	\begin{aligned}
		\sum_{\tau=0}^{\infty} \rho^{\tau}\| \mathbb{E}[\bf{u}_0^{\tau}] - \nabla  f_0(\bf{m}^{\tau})  \| <\infty,
	\end{aligned}  
\end{equation}
which indicates the convergence of $\bf{u}_0^{\tau}$. Then, according to \cite[Lemma 1]{ruszczynski1980feasible}, equation \eqref{proof_1_4} holds.

Next, according to the fact that $\bar{f}_0$ is Lipschitz continuous, it directly follows  that there exists a constant $l$ such that
\begin{equation}\label{proof_1_7}
	\begin{aligned}
		\lim_{\tau_1,\tau_2\to \infty}  \hat{f}_i^{\tau_1}(\bf{m}^{\tau_1})  - \hat{f}_i^{\tau_2}(\bf{m}^{\tau_2})  \leq  l \|   \bf{m}^{\tau_1} - \bf{m}^{\tau_2} \|.
	\end{aligned}  
\end{equation}

Finally, from \eqref{proof_1_3}, \eqref{proof_1_4} and \eqref{proof_1_7}, we can obtain that the sequences of the surrogate functions $\hat{f}_i^{\tau}(\bf{m})$ converge to $ \hat{f}_i(\bf{m}) $ almost surely.

\subsection{Proof of Theorem \ref{main_convergence}}\label{app_b}

Let $\{\bf{m}^{\tau}\}_{\tau=1}^{\infty}$ denote the sequence of model assignment policies generated by Algorithm \ref{alg_1}. According to \cite[Lemma 4]{liu2019stochastic}, we have
\begin{equation}\label{proof_2_1}
	\vspace{2.1pt}
	\begin{aligned}
		\lim_{\tau\to \infty} f_1(\bf{m}^{\tau}) \leq \bf{p}^{\textup{max}},
	\end{aligned}  
\end{equation}
\begin{equation}\label{proof_2_2}
	\begin{aligned}
		\lim_{\tau\to \infty} \| \bf{m}^{\tau} - \hat{\bf{m}}^{\tau}  \| = 0,
	\end{aligned}  
\end{equation}
where $\hat{\bf{m}}^{\tau}$ is obtained by solving problem \eqref{sca_problem} or \eqref{sca_problem_feasible}.
Then, we introduce an auxiliary variable $\tilde{\bf{m}}^{\tau}$, which is the optimal solution of the following problem,
\begin{equation}\label{proof_2_3}
	\begin{aligned}
		\tilde{\bf{m}}^{\tau}=\min_{\bf{m}} &~~ \hat{f}_0^\tau(\bf{m}) \\
		\textup{s.t.}& ~~ \hat{f}_1^\tau(\bf{m}) \leq \bf{p}^{\textup{max}} + \boldsymbol{\mu}^{\tau},\\
		&~ ~\sum_{n=1}^{N}m_n=1,\\
		& ~~0 \leq m_n \leq 1 , \forall n,
	\end{aligned}    
\end{equation}
where $\lim_{\tau\to \infty}  \boldsymbol{\mu}^{\tau} = 0$. Letting $\tau\to \infty$ in \eqref{proof_2_3} and combining \eqref{proof_1_3} and \eqref{proof_2_2} into \eqref{proof_2_3}, we have
\begin{equation}\label{proof_2_4}
	\vspace{2.1pt}
	\begin{aligned}
		\bf{m}^{*}=\min_{\bf{m}} &~~ \hat{f}_0(\bf{m}) \\
		\textup{s.t.}& ~~ \hat{f}_1(\bf{m}) \leq \bf{p}^{\textup{max}} ,\\
		&~ ~\sum_{n=1}^{N}m_n=1,\\
		& ~~0 \leq m_n \leq 1 , \forall n.
	\end{aligned}    
\end{equation}
Then, if $\bf{m}^{*}$ satisfies the Slater's condition, we have that the KKT condition of problem \eqref{proof_2_4} holds, i.e., there exists $\lambda$ such that
\vspace{2.1pt}
\begin{equation}\label{proof_2_5}
	\vspace{2.1pt}
	\begin{aligned}
		\nabla \hat{f}_0(\bf{m}^{*}) + \lambda \nabla \hat{f}_1(\bf{m}^{*}) = \bf{0}.
	\end{aligned}    
\end{equation}
Finally, it follows from Lemma 1 and \eqref{proof_2_5} that $\bf{m}^{*}$ satisfies the KKT condition of the original problem $\ca{P}_l$ as follows,
\vspace{2.1pt}
\begin{equation}\label{proof_2_6}
	\vspace{2.1pt}
	\begin{aligned}
		\nabla f_0(\bf{m}^{*}) + \lambda \nabla f_1(\bf{m}^{*}) = \bf{0}.
	\end{aligned}    
\end{equation}
This completes the proof.
%\vspace{25pt}

\bibliographystyle{ieeetr}
\bibliography{IEEEabrv,refs}

%\clearpage
% The appendix
%\input{appendix.tex}

\end{document}